\documentclass[aps,reprint,groupedaddress,nofootinbib]{revtex4-1}

\usepackage{amsmath,amssymb,graphicx,enumerate,siunitx,float}
\usepackage{hyperref}
\hypersetup{colorlinks = true, linktocpage = true, bookmarksopen = true, linkcolor = blue, urlcolor=blue, citecolor = blue}
\usepackage[margin=2cm]{geometry}
\usepackage{array}
\newcolumntype{x}[1]{>{\raggedleft\hspace{0pt}}p{#1}}

\usepackage[dvipsnames]{xcolor}

\newcommand{\revision}[1]{#1}
\newcommand{\revisiontable}[1]{#1}
\newcommand{\Deff}{D_{\mathrm{eff}}}

\newcommand{\veff}{v_{\mathrm{eff}}}

\newcommand{\Defft}{\widetilde{D}_{\mathrm{eff}}}

\graphicspath{{./Figures/}}

\begin{document}
 
\title{Advection improves homogenized models of\\ continuum diffusion in one-dimensional heterogeneous media}

\author{Elliot\hspace{0.1cm}J.\hspace{0.1cm}Carr}
\email[]{elliot.carr@qut.edu.au}
\affiliation{School of Mathematical Sciences, Queensland University of Technology (QUT), Brisbane, Australia}

\begin{abstract}
We propose an alternative method for one-dimensional continuum diffusion models with spatially variable (heterogeneous) diffusivity. Our method, which extends recent work on stochastic diffusion, assumes the constant-coefficient homogenized equation takes the form of an advection-diffusion equation with effective (diffusivity and velocity) coefficients. To calculate the effective coefficients, our approach involves solving two uncoupled boundary value problems over the heterogeneous medium and leads to coefficients depending on the spatially-varying diffusivity (as usual) as well as the boundary conditions imposed on the heterogeneous model. Computational experiments comparing our advection-diffusion homogenized model to the standard homogenized model demonstrate that including an advection term in the homogenized equation leads to improved approximations of the solution of the original heterogeneous model.
\end{abstract}


\maketitle

\section{Introduction}
Many heat and mass transport modelling problems involve heterogeneous media exhibiting spatial variability in material properties. Some examples include water and pollutant transport in groundwater aquifers composed of soils and rocks \cite{chen_2008}, heat and moisture transport within wood during drying \cite{perre_2002} and oxygen transport in biological tissues \cite{matzavinos_2016}. When such problems involve material properties that vary rapidly relative to the size of the problem domain, direct computation of the solution is expensive since one has to discretise the domain using a very fine mesh in order to capture the heterogeneity. A popular strategy for alleviating such computational issues is to formulate a simpler \emph{homogenized} model with constant coefficients that smoothes out the heterogeneity while remaining accurate \cite{chen_2008,huysmans_2007,auriault_1991,roberts_2010,davit_2013,carr_2017a,abdulle_2003,samaey_2005}. 

In this paper, we consider the one-dimensional diffusion equation in a heterogeneous medium $(0,L)$:
\begin{align}
\label{eq:diffusion_equation}
\frac{\partial u}{\partial t} = \frac{\partial}{\partial x}\left(D(x)\frac{\partial u}{\partial x}\right),
\end{align}
where $D(x) > 0$ is the spatially varying diffusivity. Our goal is to approximate the smooth or average behaviour of $u(x,t)$ by the solution of a simpler equation with spatially constant coefficients. The natural approach is to use a simpler equation of the form
\begin{align}
\label{eq:homogenized_equation}
\frac{\partial U}{\partial t} = \frac{\partial}{\partial x}\left(\Deff\frac{\partial U}{\partial x}\right),
\end{align}
where $\Deff>0$ is a constant \textit{effective}, \textit{equivalent} or \textit{homogenized} diffusivity chosen so that $U(x,t)$ provides an accurate approximation to $u(x,t)$. The standard choice for $\Deff$ is the harmonic average of $D(x)$ (see, e.g., \cite{carr_2017a,crank_1975,hornung_1997,holmes_2013,pavliotis_2008,roberts_2010,ray_2018}):
\begin{align}
\label{eq:effective_diffusivity}
\Deff = \frac{L}{\int_{0}^{L} D(x)^{-1}\,\mathrm{d}x}.
\end{align}
This definition possesses a strong theoretical foundation being the result of applying the methods of homogenization by asymptotic expansion and volume averaging \cite{carr_2017a,davit_2013,holmes_2013,pavliotis_2008} to the heterogeneous equation (\ref{eq:diffusion_equation}). The definition (\ref{eq:effective_diffusivity}) is also physically intuitive, as it can be derived by considering the boundary value problem consisting of the steady state analogue of the diffusion equation (\ref{eq:diffusion_equation}) paired with the boundary conditions $u(0) = 0$ and $u(L) = L\frac{\partial U}{\partial x}$ \cite{hornung_1997}, which impose a (constant) macroscopic gradient of $\frac{\partial U}{\partial x}$ over the medium. Solving this boundary value problem yields a solution exhibiting a constant (homogenized) flux over the medium taking the form of $q = -\Deff\frac{\partial U}{\partial x}$, where $\Deff$ is defined as in Eq (\ref{eq:effective_diffusivity}) \cite{hornung_1997}.


\begin{figure*}[!t]
\centering
\includegraphics[width=0.31\textwidth]{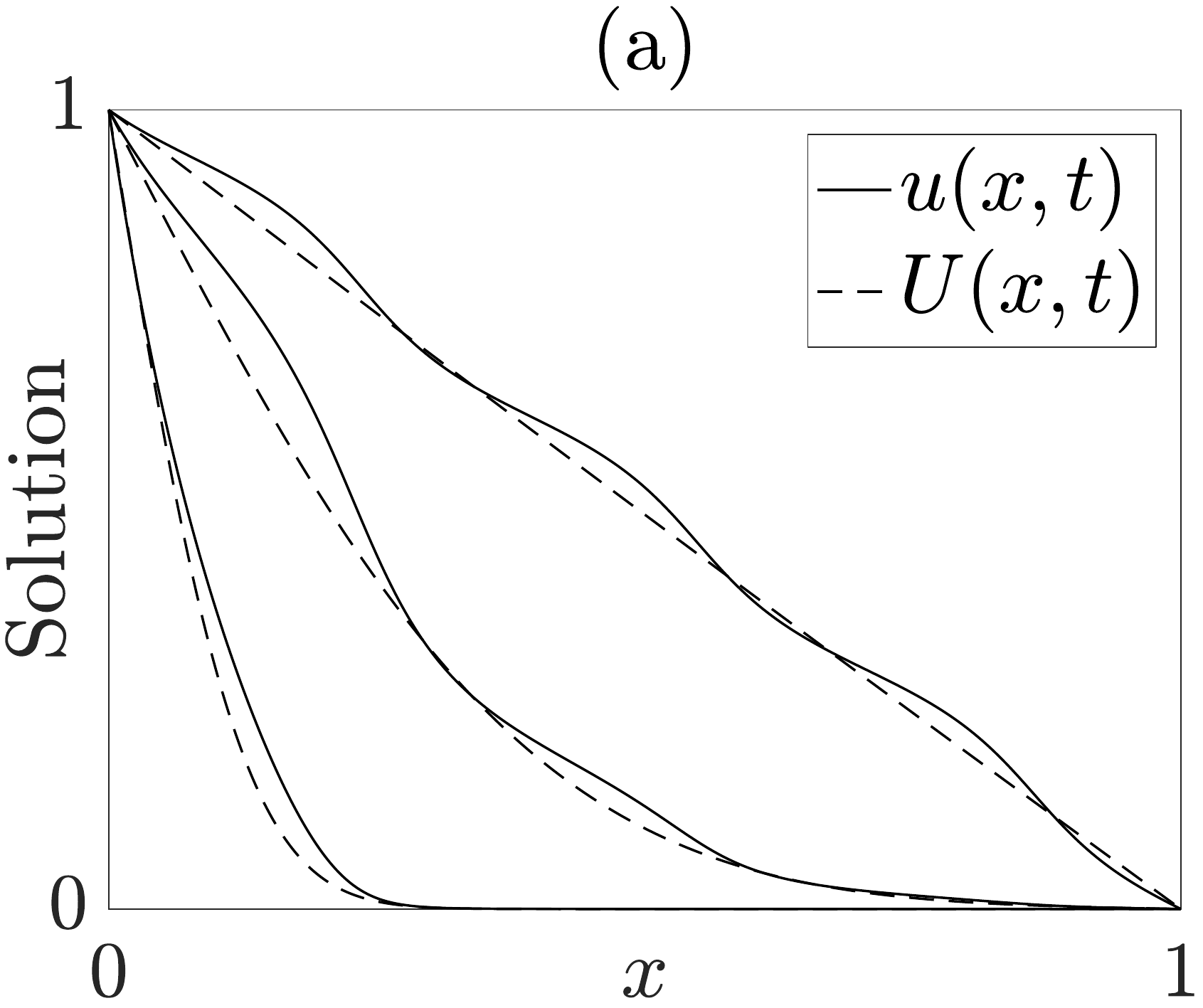}\hspace{1cm}\includegraphics[width=0.31\textwidth]{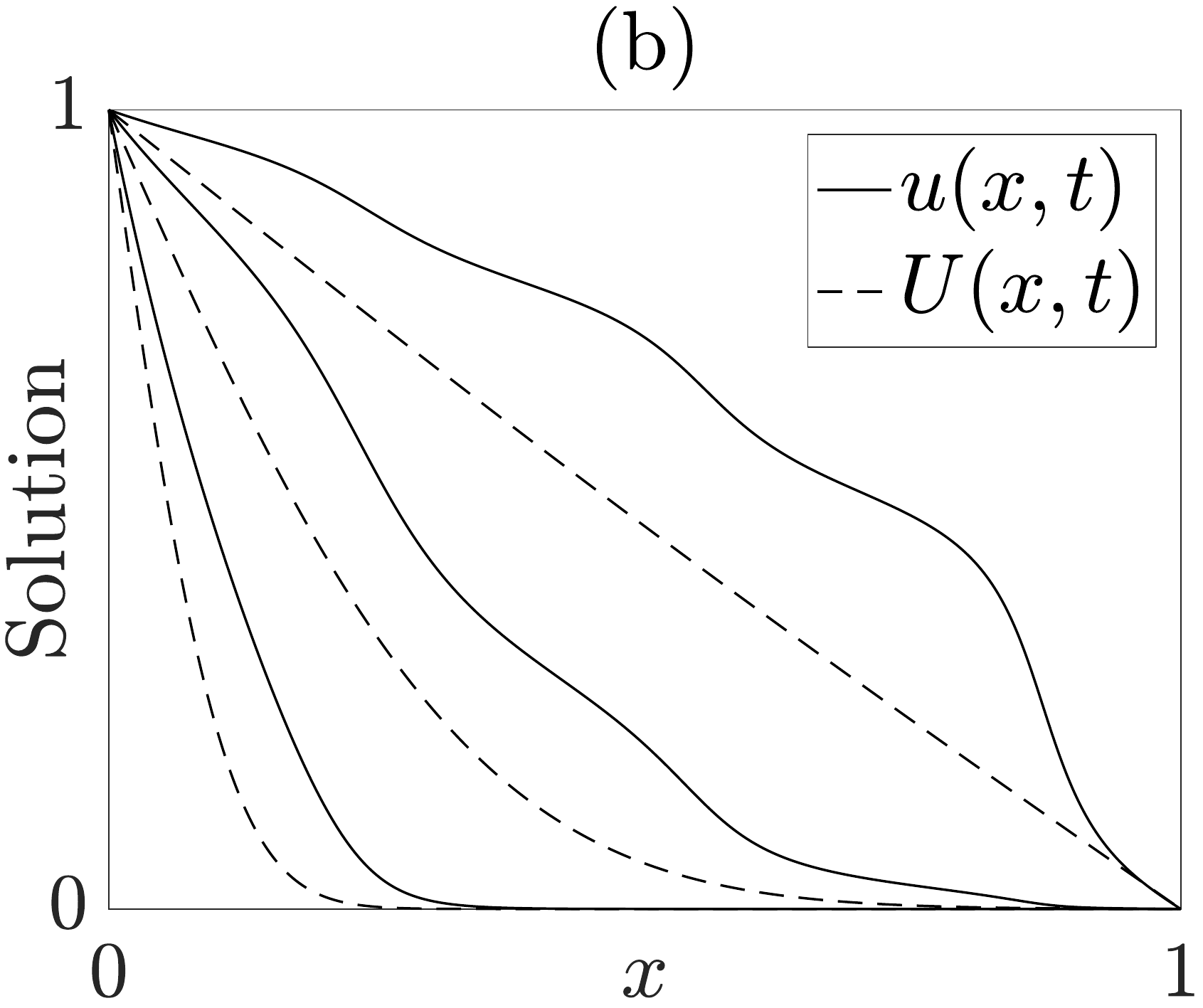}
\caption{Solution of the standard homogenized equation (\ref{eq:homogenized_equation})--(\ref{eq:effective_diffusivity}) [$U(x,t)$] benchmarked against the solution of the heterogeneous equation (\ref{eq:diffusion_equation}) [$u(x,t)$] for (a) $D(x) = 0.8 + 0.2\sin(20x)$ and (b) $D(x) = 0.8 - 0.6x + 0.2\sin(20x)$. Imposed initial and boundary conditions are $u(x,0) = U(x,0) = 0$, $u(0,t) = U(0,t) = 1$ and $u(1,t) = U(1,t) = 0$. In both plots, solutions are shown at three times $t = 10^{-2}$, $10^{-1}$, $1$.}
\label{fig:Intro}
\end{figure*}

In Figure \ref{fig:Intro}, we plot the solution of the homogenized equation (\ref{eq:homogenized_equation})--(\ref{eq:effective_diffusivity}) and the heterogeneous equation (\ref{eq:diffusion_equation}) for two choices of $D(x)$ and a particular set of initial and boundary conditions. For the first problem, the homogenized solution $U(x,t)$ provides a good approximation to the heterogeneous solution $u(x,t)$ (Figure \ref{fig:Intro}(a)). However, for the second problem, which exhibits advective behaviour in the positive $x$ direction (Figure \ref{fig:Intro}(b)), $U(x,t)$ poorly approximates $u(x,t)$  since the homogenized equation (\ref{eq:homogenized_equation}) cannot capture such behaviour. Note this behaviour becomes apparent when differentiating the diffusive flux in Eq (\ref{eq:diffusion_equation}):
\begin{align}
\label{eq:product_rule}
\frac{\partial u}{\partial t} = D(x)\frac{\partial^{2} u}{\partial x^{2}} + D'(x)\frac{\partial u}{\partial x}.
\end{align} 
To address this issue, in this paper, we present a new homogenization approach for Eq (\ref{eq:diffusion_equation}). Motivated by the results in Figure \ref{fig:Intro}, the observation (\ref{eq:product_rule}) and recent work on homogenization of random-walks through heterogeneous media \cite{carr_2019a}, our approach assumes the homogenized equation takes the form of an advection-diffusion equation:
\begin{gather}
\label{eq:fhm_pde}
\frac{\partial U}{\partial t} = \Deff\frac{\partial^{2} U}{\partial x^{2}} - \veff\frac{\partial U}{\partial x},
\end{gather}
where $\veff$ is an \textit{effective} velocity. To calculate the effective coefficients ($\Deff$ and $\veff$), our proposed method involves applying two constraints enforcing equality of appropriate measures of the spatial and temporal behaviour of the respective homogenized and heterogeneous models. This procedure requires the solution of two uncoupled boundary value problems over the heterogeneous medium and in contrast to the harmonic average definition (\ref{eq:effective_diffusivity}) leads to effective coefficients depending on the boundary conditions imposed within the heterogeneous model.

The rest of this paper is structured as follows. In section \ref{sec:heterogeneous_model}, we outline the specific heterogeneous model considered in this work including initial and boundary conditions. We then present the advection-diffusion homogenized model and describe how the effective coefficients are calculated (section \ref{sec:homogenized_model}). In section \ref{sec:computational_experiments}, computational experiments assessing the accuracy of the homogenized model are presented. Conclusions and a summary of the work are given in section \ref{sec:conclusion} along with possible avenues for future research.

\section{Heterogeneous model}
\label{sec:heterogeneous_model}
\noindent We consider the heterogeneous diffusion model:
\begin{gather}
\label{eq:hm_pde}
\frac{\partial u}{\partial t} = \frac{\partial}{\partial x}\left(D(x)\frac{\partial u}{\partial x}\right),\\
\label{eq:hm_ics}
u(x,0) = f(x),\\
\label{eq:hm_bcs}
u(0,t) = g_{0}(t),\quad u(L,t) = g_{L}(t),
\end{gather}
where $u(x,t)$ is the unknown scalar field (heterogeneous or benchmark solution), $D(x)$ is the specified spatially-varying diffusivity, $L$ is the length of the medium, $f(x)$ provides the initial solution and $g_{0}(t)$ and $g_{L}(t)$ are known functions specifying the boundary values of the solution. We make the assumption that the limits $\lim_{t\rightarrow\infty} g_{0}(t)$ and $\lim_{t\rightarrow\infty} g_{L}(t)$ exist.

\section{Homogenized model}
\label{sec:homogenized_model}
\noindent Our homogenization approach seeks to approximate the solution $u(x,t)$ of the heterogeneous model (\ref{eq:hm_pde})--(\ref{eq:hm_bcs}) by the solution $U(x,t)$ of the following advection-diffusion model with homogenized (constant) coefficients $\Deff$ and $\veff$:
\begin{gather}
\label{eq:fhm_pde}
\frac{\partial U}{\partial t} = \Deff\frac{\partial^{2} U}{\partial x^{2}} - \veff\frac{\partial U}{\partial x},\\
\label{eq:fhm_ics}
U(x,0) = f(x),\\
\label{eq:fhm_bcs}
U(0,t) = g_{0}(t),\quad U(L,t) = g_{L}(t).
\end{gather}
To determine the two unknown effective coefficients, $\Deff$ and $\veff$, we specify two constraints. First, we enforce equality of the spatial-average of the steady-state solutions of the heterogeneous (\ref{eq:hm_pde})--(\ref{eq:hm_bcs}) and homogenized (\ref{eq:fhm_pde})--(\ref{eq:fhm_bcs}) models. We express this constraint as follows:
\begin{align}
\label{eq:fhm_constraint1}
\int_{0}^{L} s(x)\,\text{d}x = \int_{0}^{L} S(x)\,\text{d}x,
\end{align} 
where $s(x)$ is the steady state solution of the heterogeneous model (\ref{eq:hm_pde})--(\ref{eq:hm_bcs}) satisfying the boundary value problem:
\begin{gather}
\label{eq:hm_pde_ss}
\frac{\mathrm{d}}{\mathrm{d}x}\left(D(x)\frac{\mathrm{d}s}{\mathrm{d}x}\right) = 0,\\
\label{eq:hm_bcs_ss}
s(0) = \lim_{t\rightarrow\infty} g_{0}(t),\quad s(L) = \lim_{t\rightarrow\infty} g_{L}(t),
\end{gather}
and $S(x)$ is the steady state solution of the homogenized model (\ref{eq:fhm_pde})--(\ref{eq:fhm_bcs}) satisfying the boundary value problem:
\begin{gather}
\label{eq:fhm_pde_ss}
\Deff\frac{\mathrm{d}^{2} S}{\mathrm{d}x^{2}} - \veff\frac{\mathrm{d}S}{\mathrm{d}x} = 0,\\
\label{eq:fhm_bcs_ss}
S(0) = \lim_{t\rightarrow\infty} g_{0}(t),\quad S(L) = \lim_{t\rightarrow\infty} g_{L}(t).
\end{gather}
The constraint (\ref{eq:fhm_constraint1}) ensures $U(x,t)$ accurately exhibits the averaged spatial behaviour of $u(x,t)$ at steady-state. To match the temporal behaviour of $U(x,t)$ and $u(x,t)$ we enforce: 
\begin{align}
\label{eq:fhm_constraint2}
\int_{0}^{L} w(x)\,\text{d}x = \int_{0}^{L} W(x)\,\text{d}x,
\end{align}
where $w(x)$ and $W(x)$ are defined as:
\begin{align}
\label{eq:wx}
w(x) &= \int_{0}^{\infty} \left[s(x) - u(x,t)\right]\,\text{d}t,\\
\label{eq:Wx}
W(x) &= \int_{0}^{\infty} \left[S(x) - U(x,t)\right]\,\text{d}t.
\end{align}
\revision{These quantities measure the (signed) distance between the steady-state and transient solutions of the heterogeneous and homogenized models over time, respectively, with the averaged values of $w(x)$ and $W(x)$ providing a simple way to quantify the temporal behaviour of each model between initial and steady state.} Attractively, both $w(x)$ and $W(x)$ can be calculated without explicit calculation of $u(x,t)$ and $U(x,t)$ appearing in the definitions \cite{carr_2019b}. Following previous work \cite{carr_2019b,carr_2018a,ellery_2012}, applying the linear operator $\mathcal{L}\varphi := \frac{\partial}{\partial x}\left(D(x)\frac{\partial\varphi}{\partial x}\right)$ to both sides of the definition (\ref{eq:wx}) and making use of the differential equations (\ref{eq:hm_pde}) and (\ref{eq:hm_pde_ss}) yields the following boundary value problem satisfied by $w(x)$:
\begin{gather}
\label{eq:hm_pde_m}
\frac{\mathrm{d}}{\mathrm{d}x}\left(D(x)\frac{\mathrm{d}w}{\mathrm{d}x}\right) = r(x),\\
\label{eq:hm_bcs_m1}
w(0) = \int_{0}^{\infty}\left[s(0) - g_{0}(t)\right]\,\mathrm{d}t,\\ 
\label{eq:hm_bcs_m2}
w(L) = \int_{0}^{\infty}\left[s(L) - g_{L}(t)\right]\,\mathrm{d}t,
\end{gather}
where $r(x) = f(x) - s(x)$. We remark that the boundary conditions (\ref{eq:hm_bcs_m1})--(\ref{eq:hm_bcs_m2}) follow directly from the definition (\ref{eq:wx}) and the heterogeneous model boundary conditions (\ref{eq:hm_bcs}) with the imposed values of $w(0)$ and $w(L)$ calculated by directly evaluating the integrals since the values of $s(0)$ and $s(L)$ (\ref{eq:hm_bcs_ss}) and boundary functions $g_{0}(t)$ and $g_{L}(t)$ are known. In a similar manner to that carried out for $w(x)$, except now with the linear operator defined by $\mathcal{L}\varphi := \Deff\frac{\partial^{2}\varphi}{\partial x} - \veff\frac{\partial\varphi}{\partial x}$, the following boundary value problem is derived for $W(x)$:
\begin{gather}
\label{eq:hm_pde_M}
\Deff\frac{\mathrm{d}^{2}W}{\mathrm{d}x^{2}} - \veff\frac{\mathrm{d}W}{\mathrm{d}x} = R(x),\\
\label{eq:hm_bcs_M1}
W(0) = \int_{0}^{\infty}\left[S(0) - g_{0}(t)\right]\,\mathrm{d}t,\\ 
\label{eq:hm_bcs_M2}
W(L) = \int_{0}^{\infty}\left[S(L) - g_{L}(t)\right]\,\mathrm{d}t,
\end{gather}
where $R(x) = f(x)-S(x)$. 

Since the solutions $S(x)$ and $W(x)$ will depend nonlinearly on $\Deff$ and $\veff$, the two constraints (\ref{eq:fhm_constraint1}) and (\ref{eq:fhm_constraint2}) together define a pair of coupled nonlinear equations:
\begin{align}
\label{eq:fhm_nonlinear_system}
\mathbf{F}(\mathbf{c}) = \left(F_{1}(\mathbf{c}),F_{2}(\mathbf{c})\right)^{T} = \mathbf{0},
\end{align}
whose solution $\mathbf{c} = \left[\Deff,\veff\right]^{T}$ provides the effective coefficients supplied to the homogenized model (\ref{eq:fhm_pde})--(\ref{eq:fhm_bcs}). The form of the component functions, $F_{1}$ and $F_{2}$, are formulated by solving the various boundary value problems numerically. In this work, we employ a vertex-centered finite volume method on a uniform grid consisting of $N_{x}$ nodes with node spacing $h = L/(N_{x}-1)$. Let $s_{k}$, $S_{k}$, $w_{k}$ and $W_{k}$ denote the numerical approximation to the solutions of the boundary value problems (\ref{eq:hm_pde_ss})--(\ref{eq:hm_bcs_ss}), (\ref{eq:fhm_pde_ss})--(\ref{eq:fhm_bcs_ss}), (\ref{eq:hm_pde_m})--(\ref{eq:hm_bcs_m2}) and (\ref{eq:hm_pde_M})--(\ref{eq:hm_bcs_M2}) at $x = x_{k} := (k-1)h$ for $k = 1,\hdots,N_{x}$. Using these solutions, a simple trapezoidal rule is applied to evaluate the integrals in the constraint equations (\ref{eq:fhm_constraint1}) and (\ref{eq:fhm_constraint2}) yielding the component functions:
\begin{align}
\label{eq:F1}
\hspace*{-0.05cm}F_{1}(\mathbf{c}) &= \frac{h}{2}\sum_{k=2}^{N_{x}} \left[S_{k-1} + S_{k}\right] - \frac{h}{2}\sum_{k=2}^{N_{x}} \left[s_{k-1} + s_{k}\right],\\
\label{eq:F2}
\hspace*{-0.05cm}F_{2}(\mathbf{c}) &= \frac{h}{2}\sum_{k=2}^{N_{x}} \left[W_{k-1} + W_{k}\right] - \frac{h}{2}\sum_{k=2}^{N_{x}} \left[w_{k-1} + w_{k}\right].
\end{align}
Note the values of $S_{k}$ and $W_{k}$ depend nonlinearly on $\Deff$ and $\veff$ so $F_{1}$ and $F_{2}$ are nonlinear functions of $\mathbf{c}$.

\section{Computational experiments}
\label{sec:computational_experiments}
We now compare the solution of the advection-diffusion homogenized model (\ref{eq:fhm_pde})--(\ref{eq:fhm_bcs}), $U(x,t)$, to the benchmark solution, $u(x,t)$, of the heterogeneous model (\ref{eq:hm_pde})--(\ref{eq:hm_bcs}). We also compare these solutions to the solution of the standard diffusion-only homogenized model, $\widetilde{U}(x,t)$, where the effective equation takes the form of the diffusion equation with constant harmonic-averaged effective diffusivity:
\begin{gather}
\label{eq:cfhm_pde}
\frac{\partial \widetilde{U}}{\partial t} = \Defft\frac{\partial^{2} \widetilde{U}}{\partial x^{2}},\\
\widetilde{U}(x,0) = f(x),\\ 
\widetilde{U}(0,t) = g_{0}(t),\quad \widetilde{U}(L,t) = g_{L}(t),\\
\label{eq:cfhm_De}
\text{with $\displaystyle\Defft = \frac{L}{\int_{0}^{L} D(x)^{-1}\,\mathrm{d}x}$}.
\end{gather}

\begin{table*}[!tb]
\setlength{\tabcolsep}{10pt}
\renewcommand{\arraystretch}{1.5}
\begin{tabular}{|c|p{0.85\textwidth}|}
\hline
Case & Description\\
\hline
A & $D(x) = 0.5 + 0.2\sin(20x)$, $f(x) = 0$, $g_{0}(t) = 1$ and $g_{L}(t) = 0$.\\
B & $D(x) = 0.8 - 0.6x + 0.2\sin(20x)$, $f(x) = 0$, $g_{0}(t) = 1$ and $g_{L}(t) = 0$.\\
C & $D(x) = 0.5 + 0.2\sin(x/\varepsilon)$ with $\varepsilon = 0.005$, $f(x) = 0$, $g_{0}(t) = 1$ and $g_{L}(t) = 0$.\\
\revisiontable{D} & \revisiontable{$D(x) = 0.1\exp(2.2x^{2}) + 0.05\sin(80x)$, $f(x) = 0$, $g_{0}(t) = 1$ and $g_{L}(t) = 0$.}\\
\revisiontable{E} & \revisiontable{$D(x) = 0.5 + 0.5\exp(x-1)\cos(80(1-x))$, $f(x) = 0$, $g_{0}(t) = 1$ and $g_{L}(t) = 0$.}\\
\revisiontable{F} & \revisiontable{$D(x)$ is piecewise linear on the sub-intervals $((i-1)H,iH)$ where $i = 1,\hdots,50$ and $H = 1/50$. At locations $x_{i} = iH$ for $i = 0,\hdots,50$, $D(x_{i}) = 0.9 - 0.8 / (1 + \exp(20(x-0.5))) + \varepsilon$, where $\varepsilon$ is a random number generated from a normal distribution with support $[-0.05,0.05]$. For all $i = 1,\hdots,50$, within sub-interval $((i-1)H,iH)$ the linear form of $D(x)$ is constructed to interpolate the values previously assigned at $x_{i-1} = (i-1)H$ and $x_{i} = iH$. $f(x) = 0$, $g_{0}(t) = 1$ and $g_{L}(t) = 0$.}\\
G & $D(x) = 0.5 + 0.24(\sin(20x) + \sin(80x))$, $f(x) = 0$, $g_{0}(t) = 1$ and $g_{L}(t) = 0.75(1-\exp(-25t))$.\\
H & $D(x)$ is piecewise constant on the sub-intervals $((i-1)H,iH)$ where $i = 1,\hdots,16$ and $H = 1/16$. Within each sub-interval the constant value of $D(x)$ is assigned randomly from a uniform distribution with support $[0.01,0.99]$. $f(x) = \exp(-30(x-0.5)^2)$, $g_{0}(t) = 0$ and $g_{L}(t) = 10^{-6}$.\\
I & $D(x)$ is piecewise linear on the sub-intervals $((i-1)H,iH)$ where $i = 1,\hdots,24$ and $H = 1/24$. At locations $x_{i} = iH$ for $i = 0,\hdots,24$, we assign a random value of the diffusivity generated from a uniform distribution with support $[0.01,0.99]$. For all $i = 1,\hdots,24$, within sub-interval $((i-1)H,iH)$ the linear form of $D(x)$ is constructed to interpolate the random values previously assigned at $x_{i-1} = (i-1)H$ and $x_{i} = iH$. $f(x) = 2x$ if $x\in[0,0.5]$ otherwise $f(x) = 2-2x$ if $x\in[0.5,1]$, $g_{0}(t) = 0$ and $g_{L}(t) = 0.5$.\\
\hline
\end{tabular}
\caption{Problem descriptions for the nine test cases used in the computational experiments of Section \ref{sec:computational_experiments}. For each test case, this table identifies  the spatially-variable diffusivity $D(x)$, initial solution $f(x)$, and boundary values $g_{0}(t)$ and $g_{L}(t)$ appearing in the heterogeneous model (\ref{eq:hm_pde})--(\ref{eq:hm_bcs}).}
\label{tab:test_cases}
\end{table*} 

\begin{table*}[]
\centering
\setlength{\tabcolsep}{10pt}
\renewcommand{\arraystretch}{1.5}
\begin{tabular}{|c|rrr|rr|}
\hline
& \multicolumn{3}{c|}{Homogenized advection-diffusion model} & \multicolumn{2}{c|}{Homogenized diffusion-only model}\\
\hline
Case & $\Deff$ & $\veff$ & \text{Error} & $\Defft$ & \text{Error}\\
\hline
A & \num{0.497} & \num{0.079} & \num{1.48e-02} & \num{0.464} & \num{1.81e-02}\\
B & \num{0.409} & \num{0.805} & \num{2.36e-02} & \num{0.345} & \num{1.38e-01}\\
C & \num{0.462} & \num{0.009} & \num{1.51e-03} & \num{0.459} & \num{1.91e-03}\\
\revisiontable{D} & \revisiontable{\num{0.160}} & \revisiontable{\num{-0.319}} & \revisiontable{\num{1.36e-02}} & \revisiontable{\num{0.162}} & \revisiontable{\num{7.22e-02}}\\
\revisiontable{E} & \revisiontable{\num{0.318}} & \revisiontable{\num{0.374}} & \revisiontable{\num{1.91e-02}} & \revisiontable{\num{0.324}} & \revisiontable{\num{6.57e-02}}\\
\revisiontable{F} & \revisiontable{\num{0.186}} & \revisiontable{\num{-0.622}} & \revisiontable{\num{2.06e-02}} & \revisiontable{\num{0.193}} & \revisiontable{\num{1.20e-01}}\\
G & \num{0.342} & \num{0.097} & \num{2.93e-02} & \num{0.306} & \num{3.18e-02}\\
H & \num{0.312} & \num{-0.317} & \num{1.69e-02} & \num{0.215} & \num{4.32e-02}\\
I & \num{0.427} & \num{0.062} & \num{1.51e-02} & \num{0.381} & \num{1.87e-02}\\
\hline
\end{tabular}
\caption{Mean absolute errors (\ref{eq:epsilon})--(\ref{eq:epsilon_tilde}) and effective coefficients for the new homogenized model (\ref{eq:fhm_pde})--(\ref{eq:fhm_bcs}) and standard homogenized model (\ref{eq:cfhm_pde})--(\ref{eq:cfhm_De}).}
\label{tab:errors}
\end{table*}

To obtain the the effective coefficients $\Deff$ and $\veff$ for the advection-diffusion homogenized model (\ref{eq:fhm_pde})--(\ref{eq:fhm_bcs}), we solve the nonlinear system (\ref{eq:fhm_nonlinear_system})--(\ref{eq:F2}) using MATLAB's in-built \verb"fsolve" function. To solve the heterogeneous model (\ref{eq:hm_pde})--(\ref{eq:hm_bcs}), the advection-diffusion homogenized  model (\ref{eq:fhm_pde})--(\ref{eq:fhm_bcs}) and the diffusion-only homogenized model (\ref{eq:cfhm_pde})--(\ref{eq:cfhm_De}), we use a vertex-centered finite volume method on a uniform grid consisting of $N_{x}$ nodes with node spacing $h = L/(N_{x}-1)$ (as in Section \ref{sec:homogenized_model}). The resulting system of differential equations is then solved numerically using MATLAB's in-built \verb"ode15s" function.  Further implementation details can be found in our code available on GitHub: \href{https://github.com/elliotcarr/Carr2019b}{https://github.com/elliotcarr/Carr2019b}. Let $u_{k}^{j}$, $U_{k}^{j}$ and $\widetilde{U}_{k}^{j}$ denote the resulting numerical approximations to $u(x,t)$, $U(x,t)$ and $\widetilde{U}(x,t)$ at $x = x_{k} := (k-1)h$ for $k = 1,\hdots,N_{x}$ and $t = t_{j} := j\tau$ for $j = 1,\hdots,N_{t}$, where $\tau > 0$ and $N_{t}\in\mathbb{N}^{+}$. Using these solutions we define the following mean absolute errors:
\begin{align}
\label{eq:epsilon}
\text{Error} = \frac{1}{N_{t}N_{x}}\sum_{j=1}^{N_{t}}\sum_{k=1}^{N_{x}} \bigl|U_{k}^{j} - u_{k}^{j}\bigr|,
\end{align}
for the the advection-diffusion homogenized model (\ref{eq:fhm_pde})--(\ref{eq:fhm_bcs}) and
\begin{align}
\label{eq:epsilon_tilde}
\text{Error} = \frac{1}{N_{t}N_{x}}\sum_{j=1}^{N_{t}}\sum_{k=1}^{N_{x}} \bigl|\widetilde{U}_{k}^{j} - u_{k}^{j}\bigr|,
\end{align}
for the diffusion-only homogenized model (\ref{eq:cfhm_pde})--(\ref{eq:cfhm_De}). 

\begin{figure*}[p]
\centering
\def\figurewidth{0.23\textwidth}
\includegraphics[width=\figurewidth]{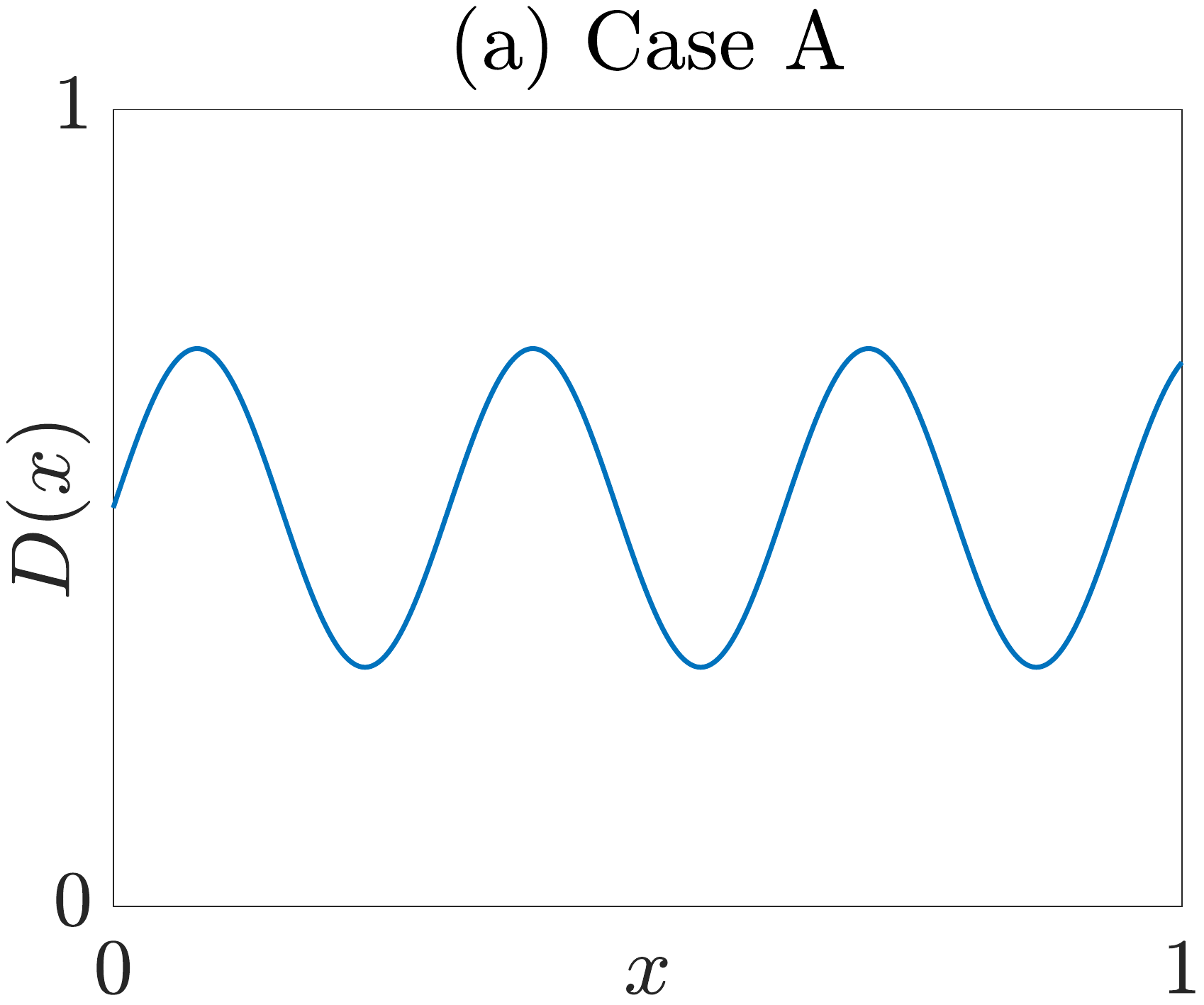}\hspace{0.35cm}\includegraphics[width=\figurewidth]{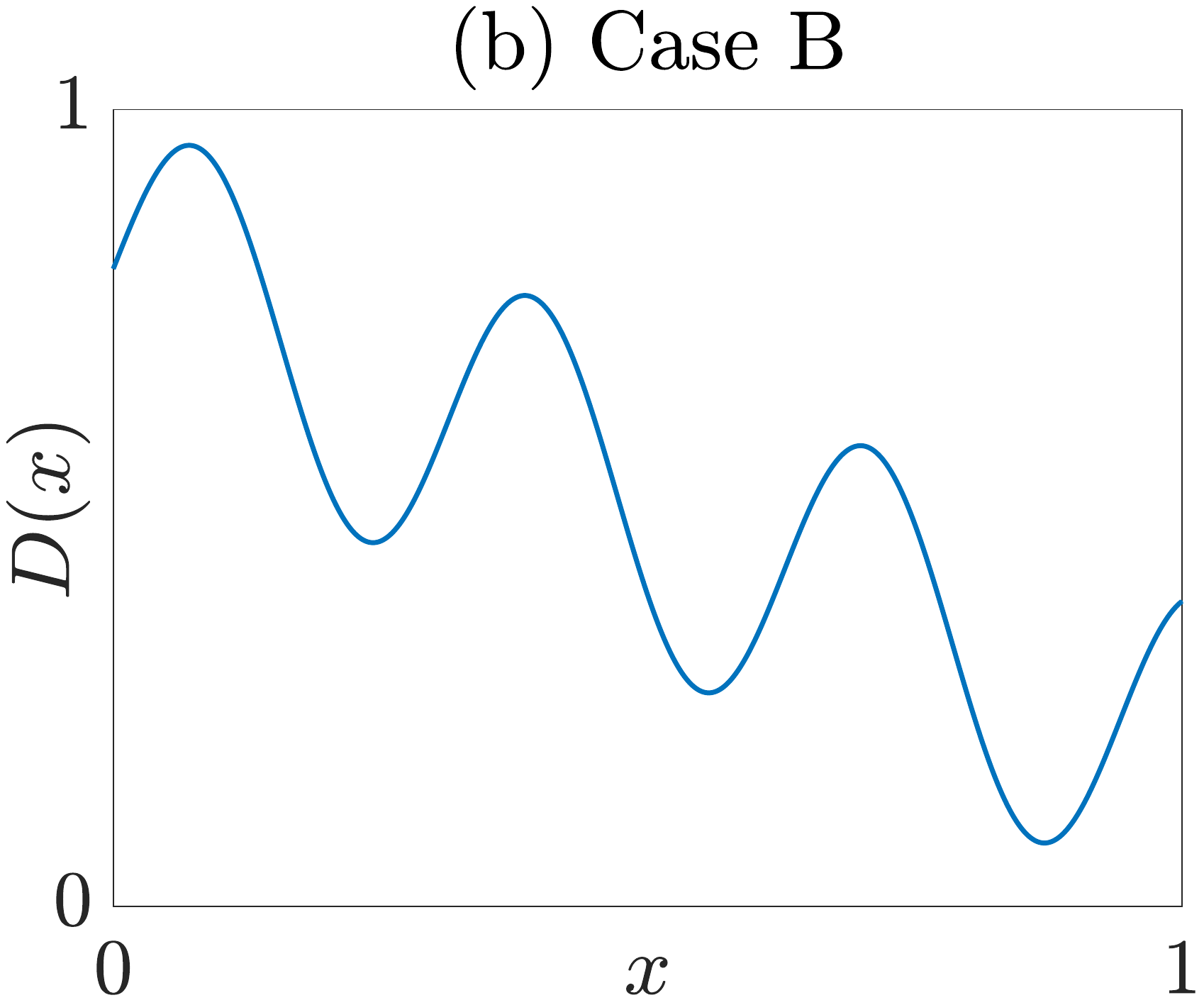}\hspace{0.25cm}
\includegraphics[width=\figurewidth]{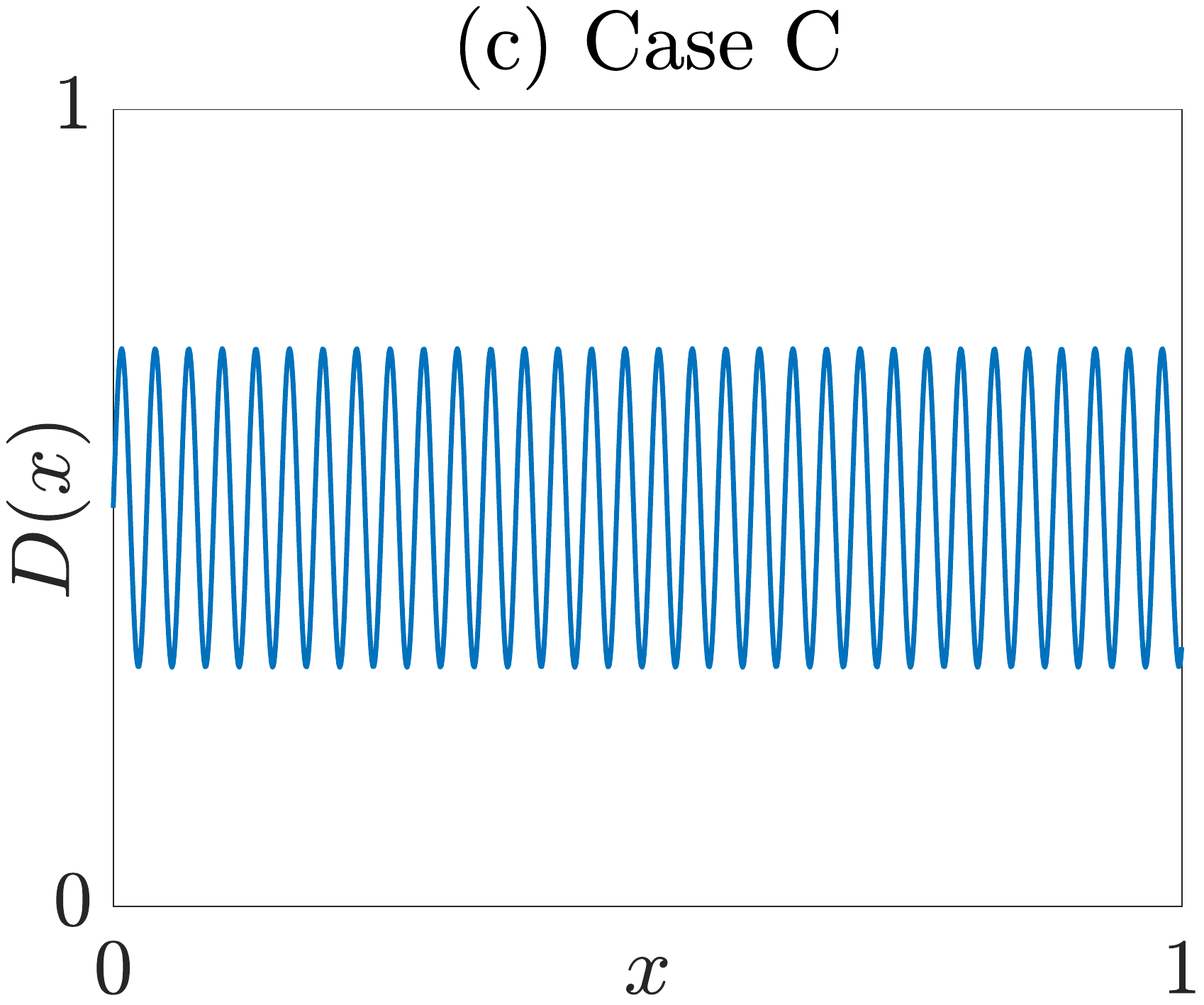}\\\includegraphics[width=\figurewidth]{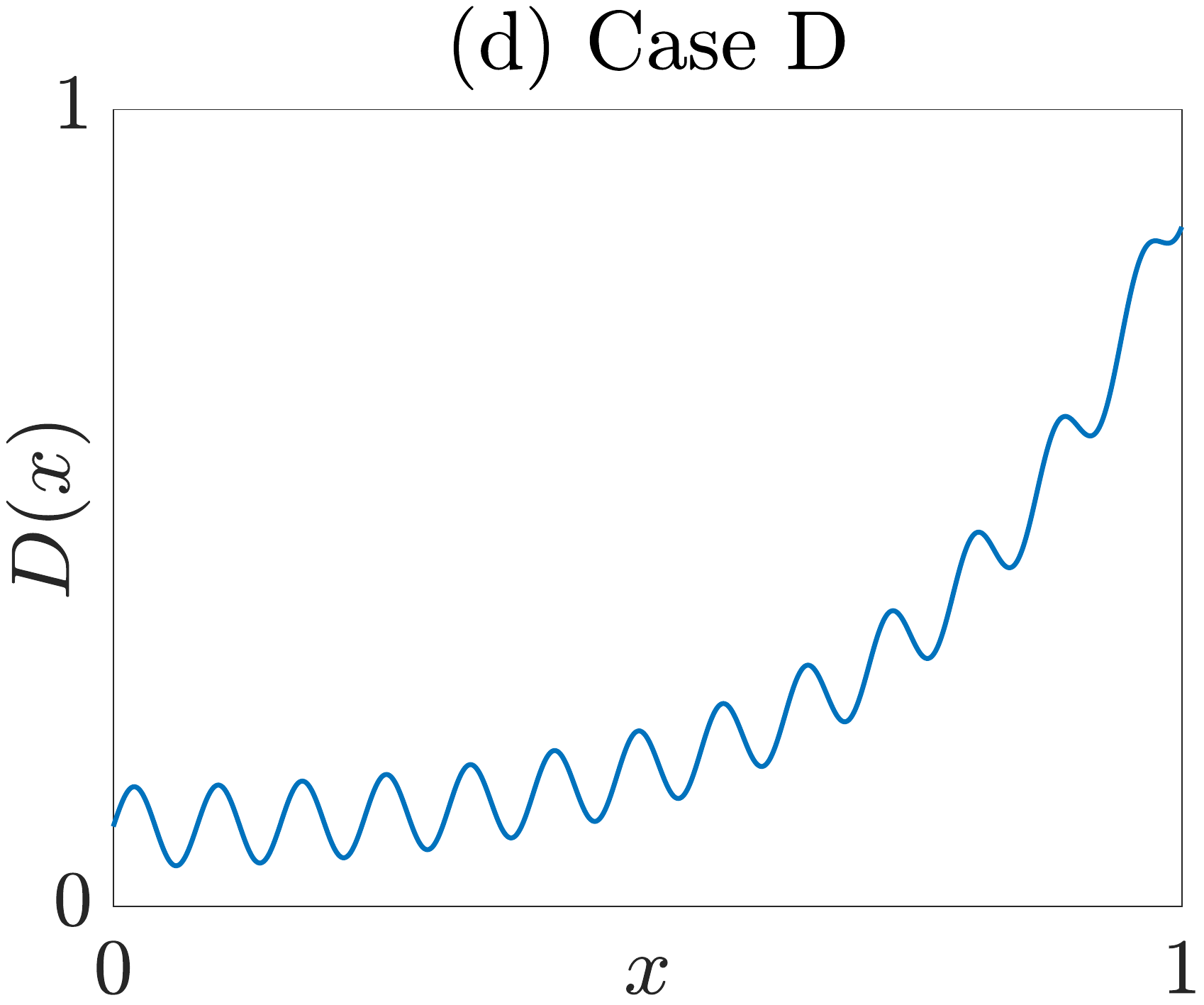}\hspace{0.35cm}\includegraphics[width=\figurewidth]{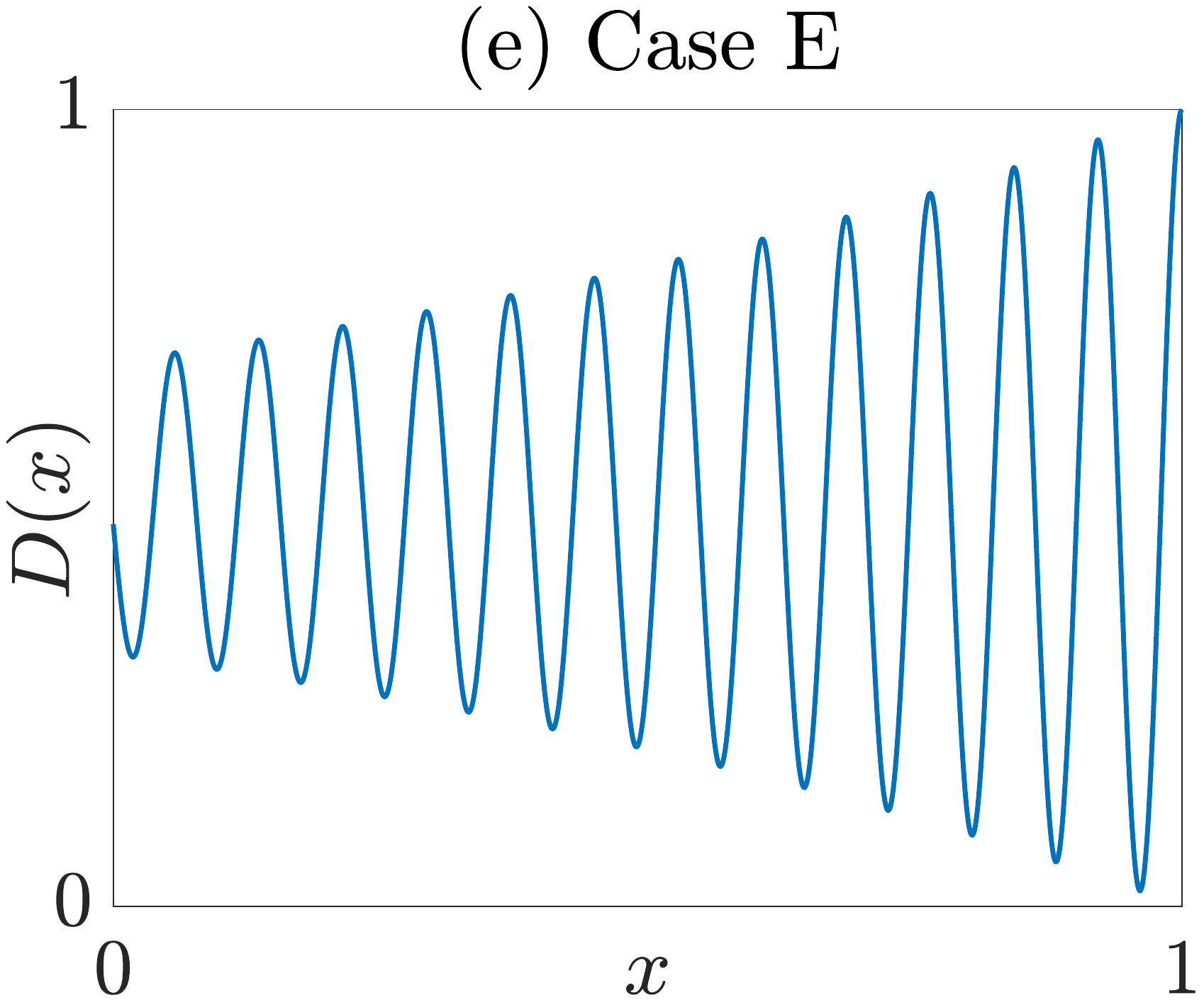}\hspace{0.25cm}
\includegraphics[width=\figurewidth]{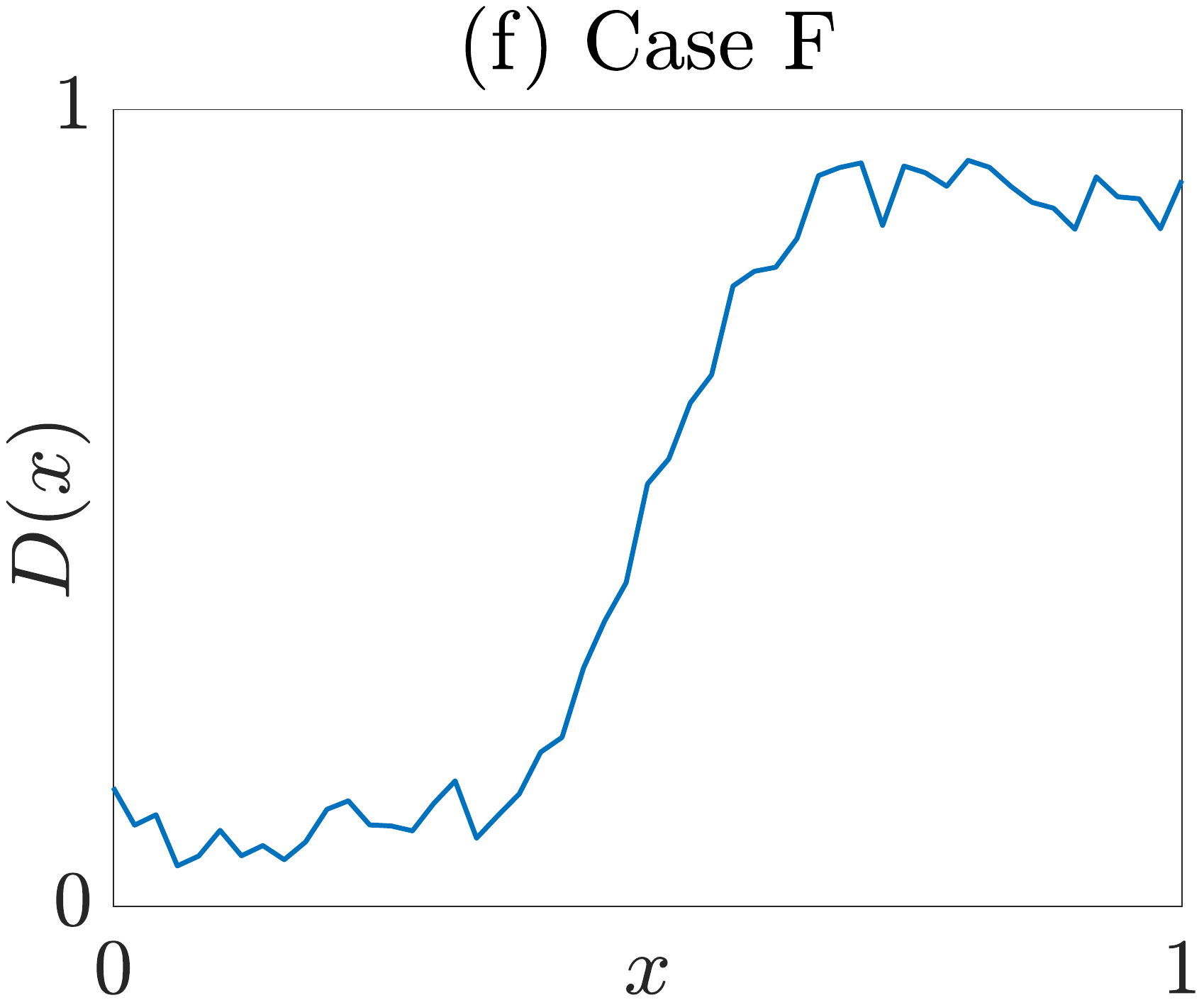}\\
\includegraphics[width=\figurewidth]{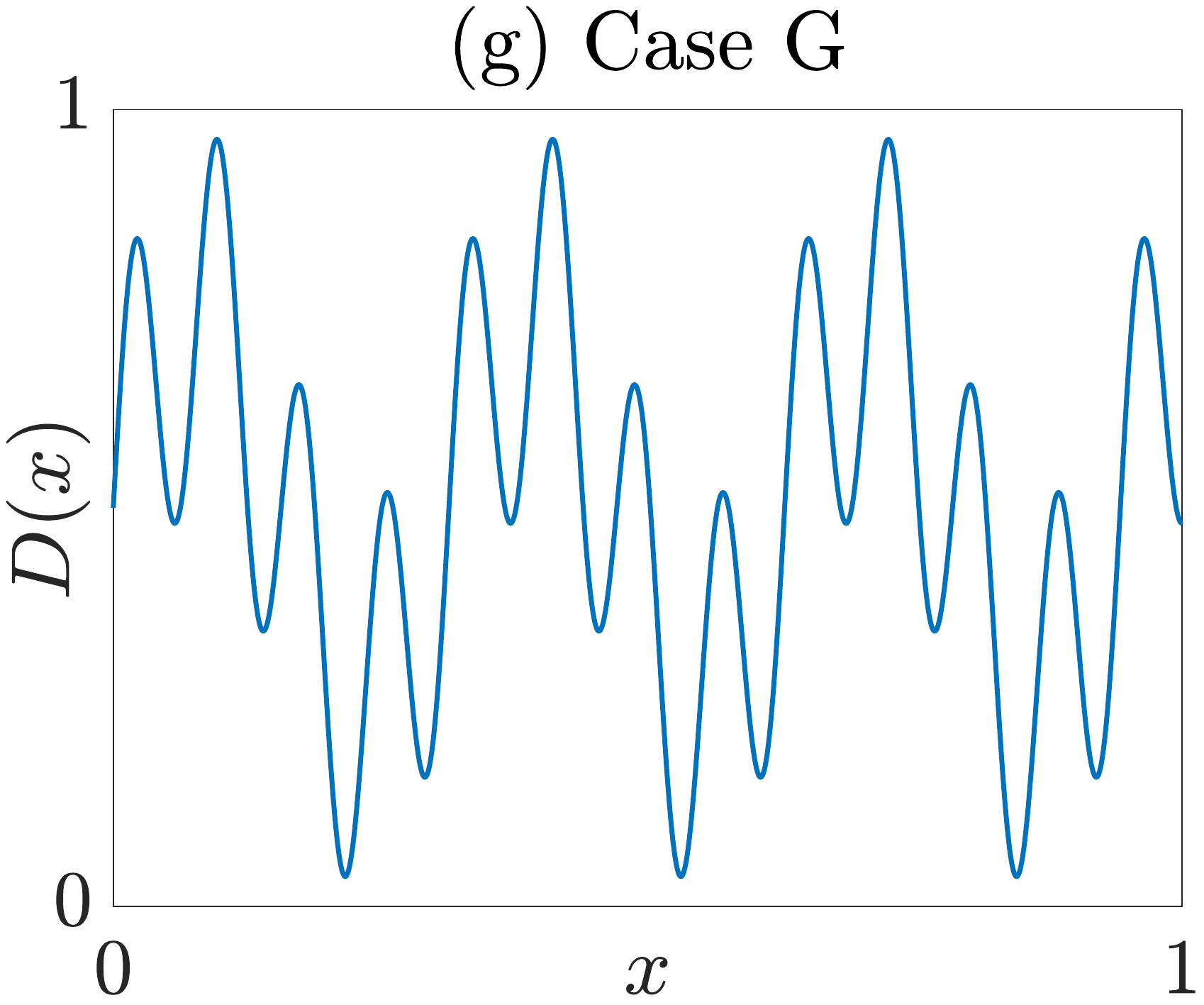}\hspace{0.35cm}\includegraphics[width=\figurewidth]{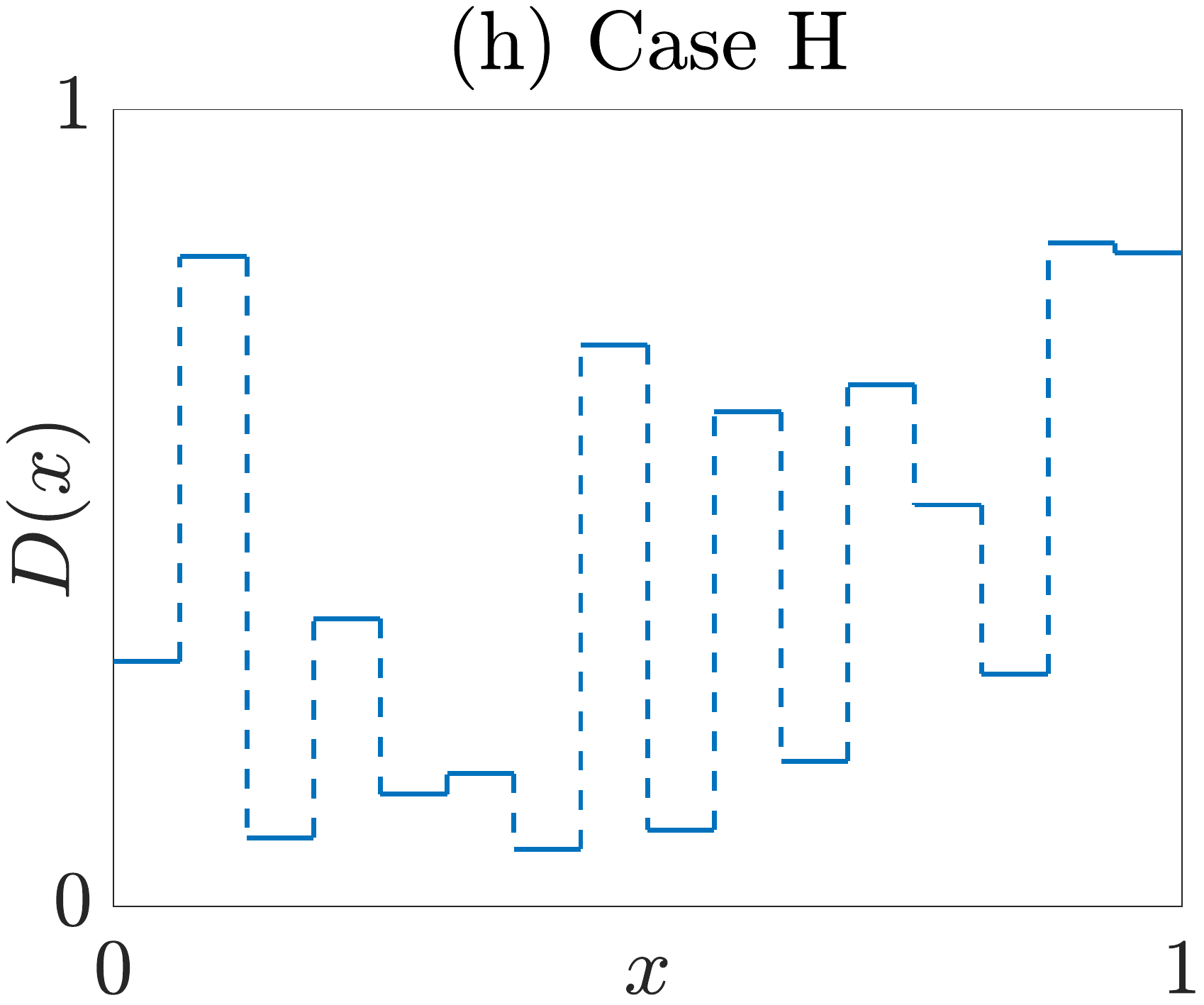}\hspace{0.25cm}
\includegraphics[width=\figurewidth]{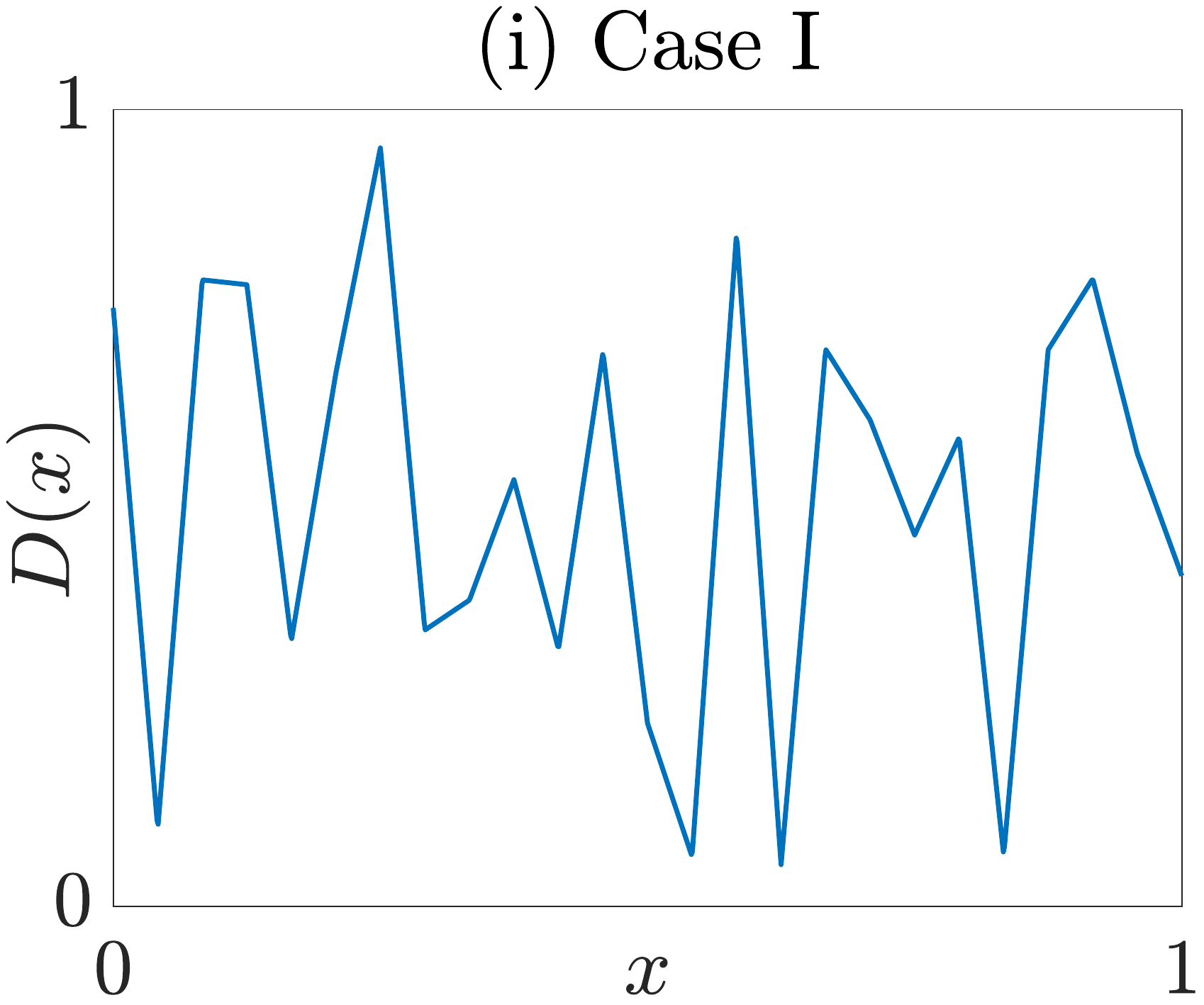}\\[0.8cm]
\includegraphics[width=0.2\textwidth]{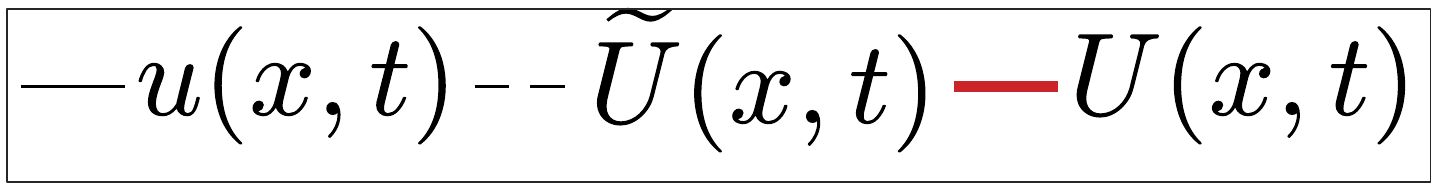}\\[0.1cm]
\includegraphics[width=\figurewidth]{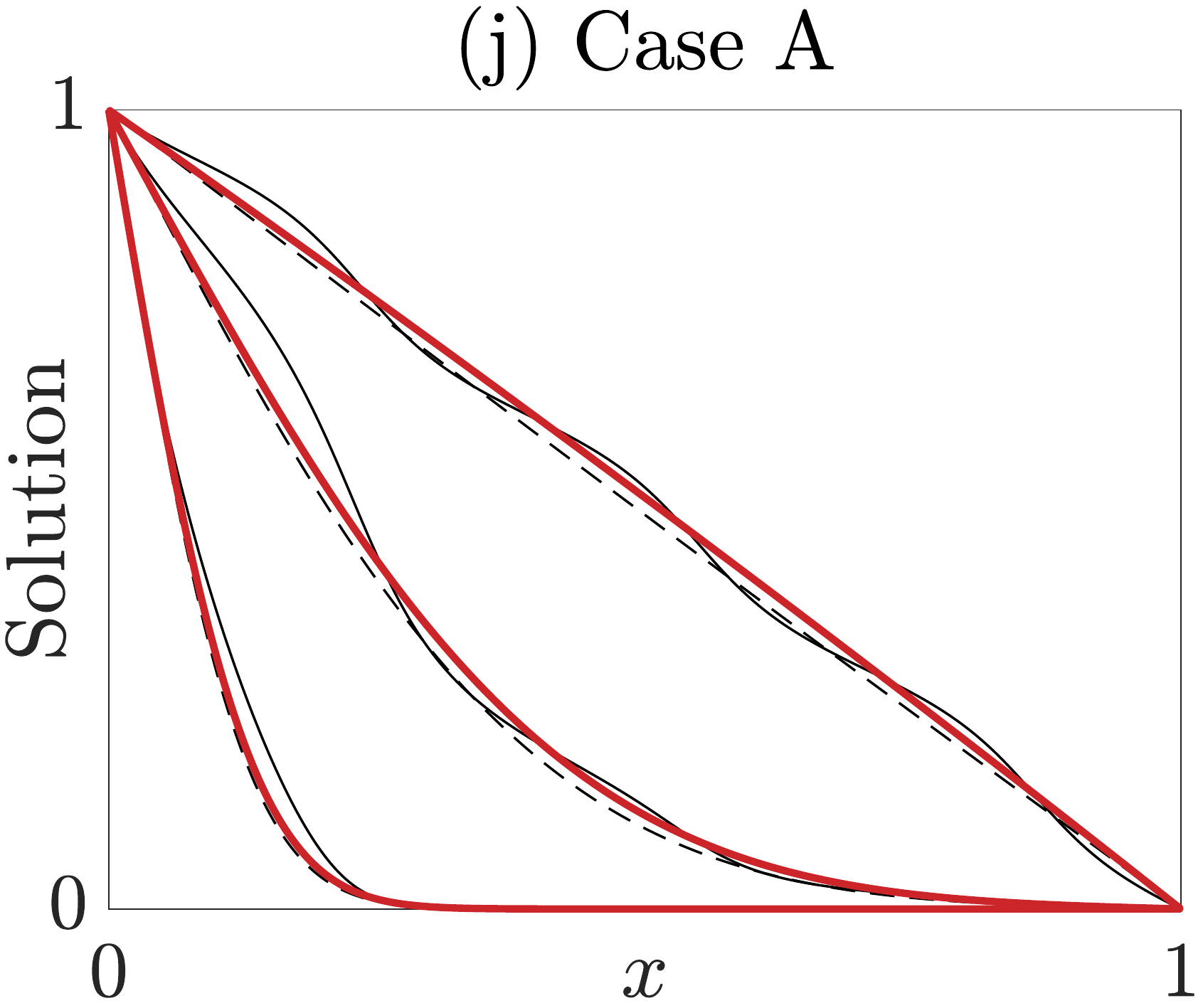}\hspace{0.35cm}\includegraphics[width=\figurewidth]{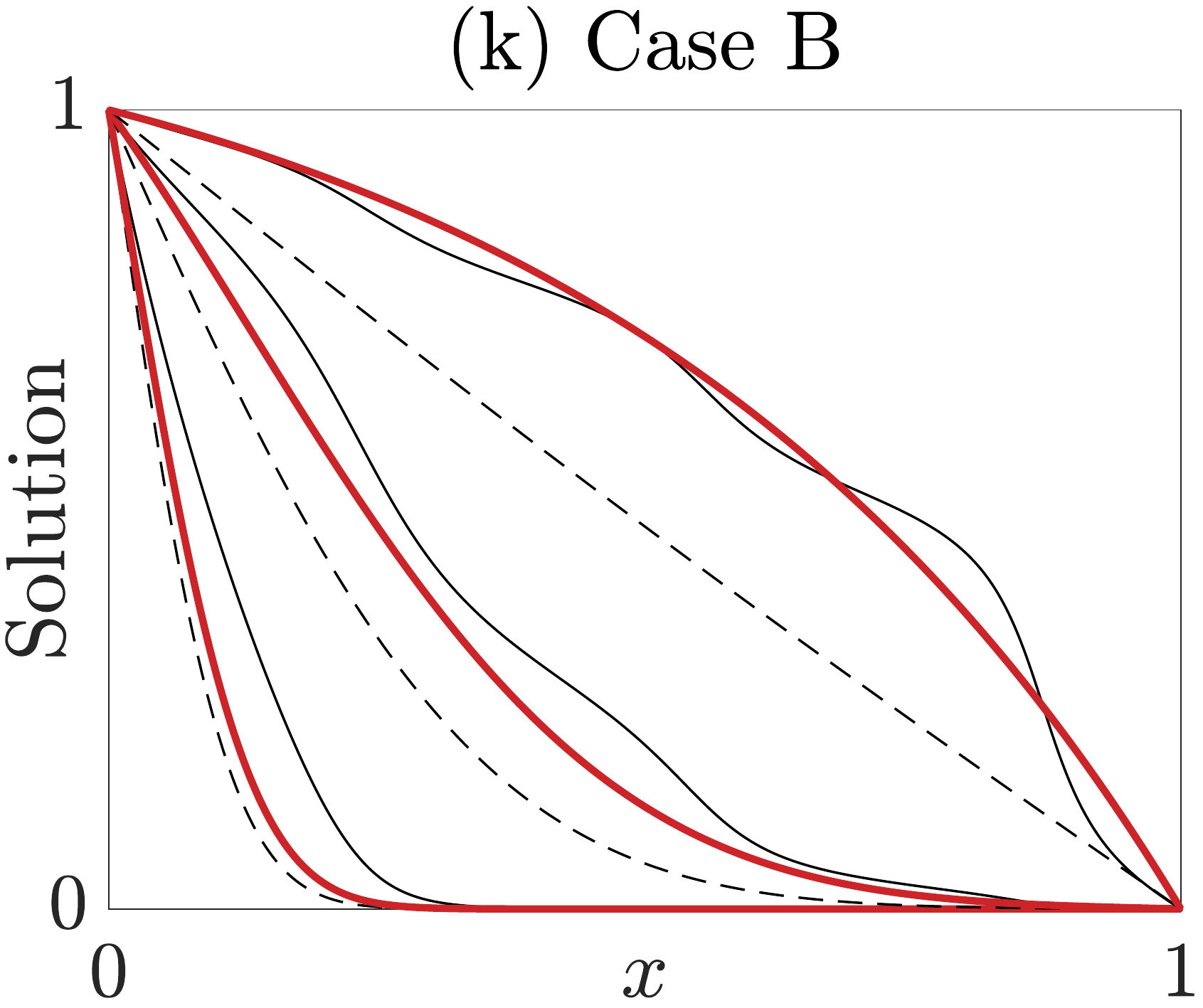}\hspace{0.25cm}
\includegraphics[width=\figurewidth]{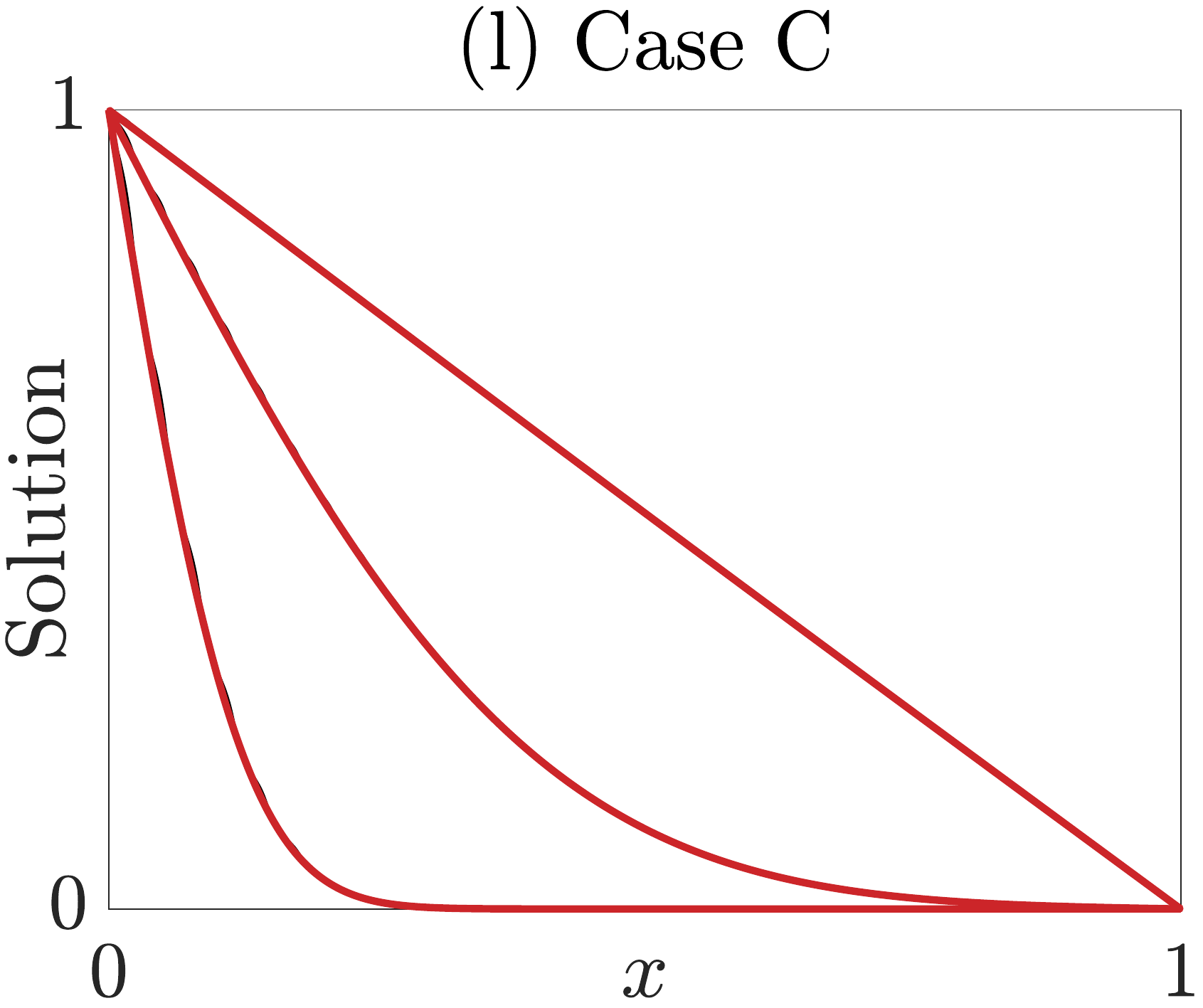}\\
\includegraphics[width=\figurewidth]{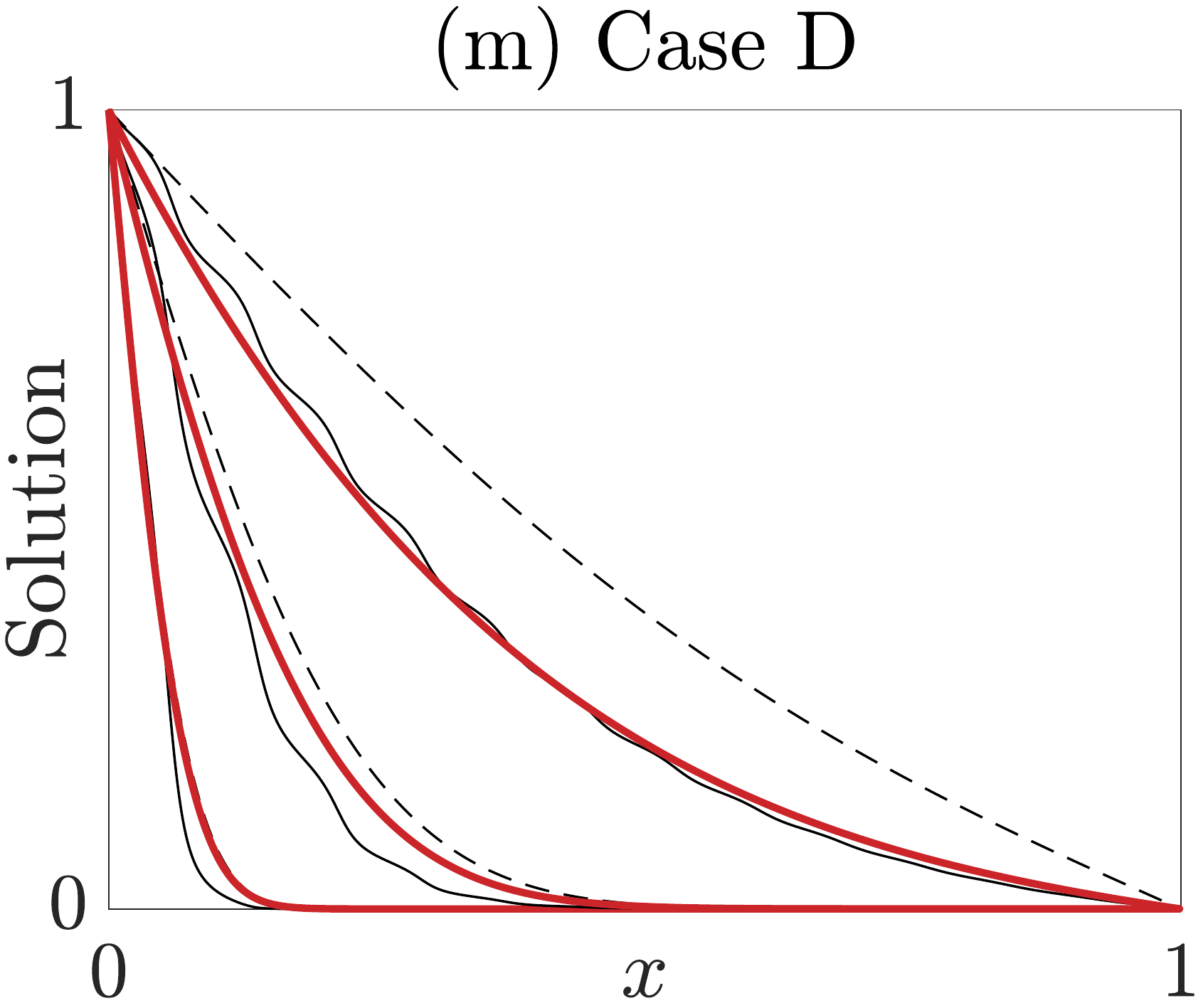}\hspace{0.35cm}\includegraphics[width=\figurewidth]{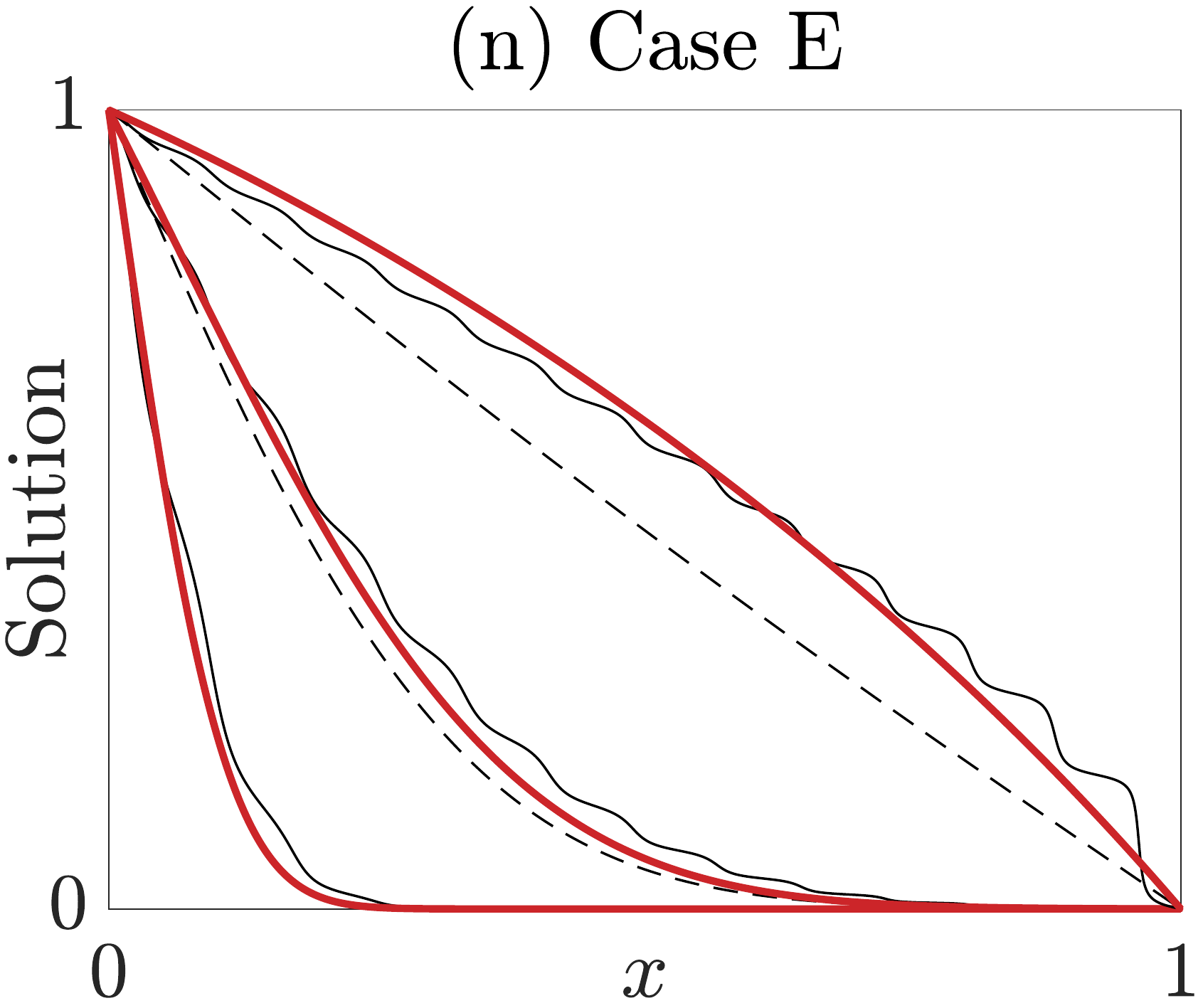}\hspace{0.25cm}
\includegraphics[width=\figurewidth]{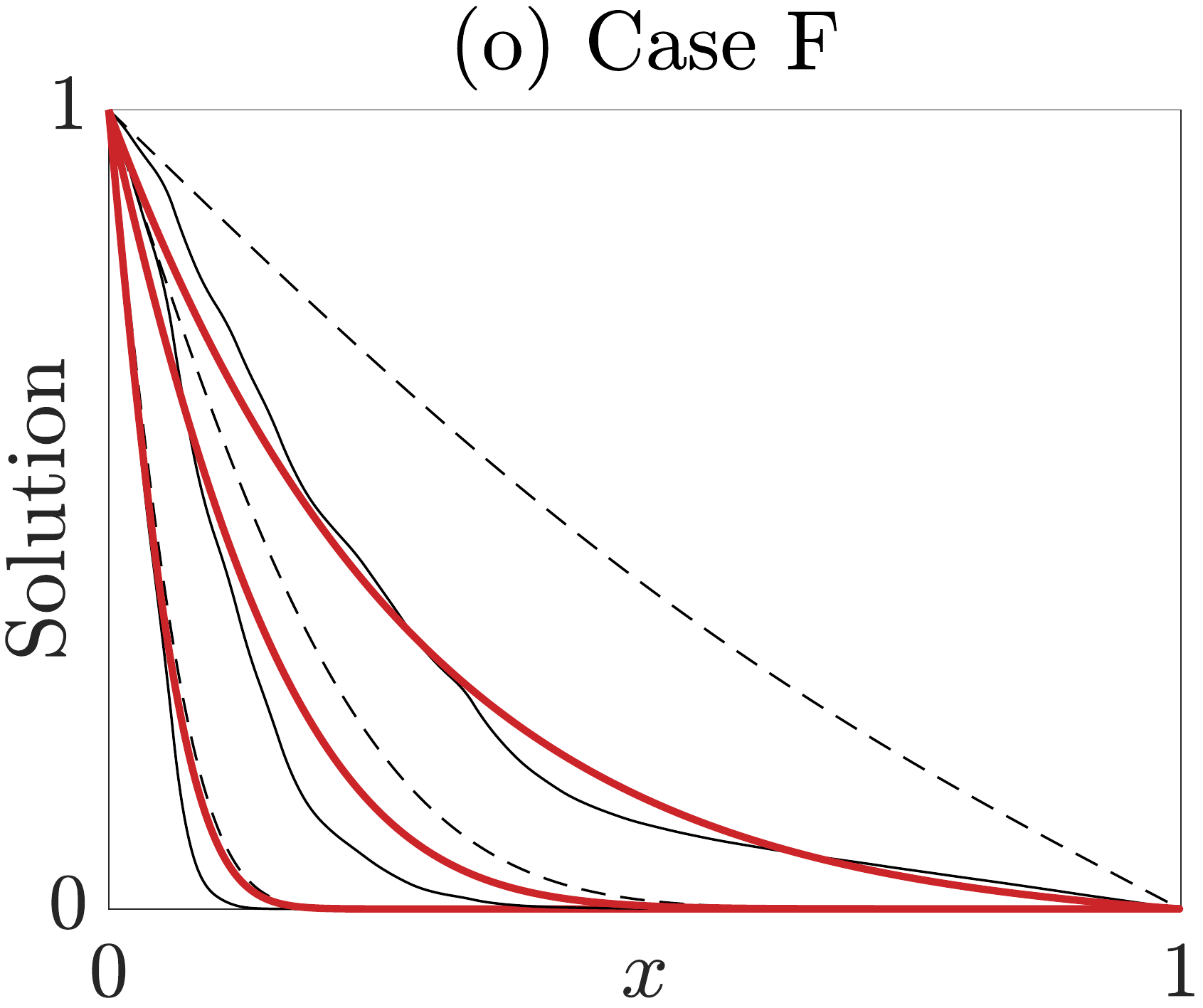}\\
\includegraphics[width=\figurewidth]{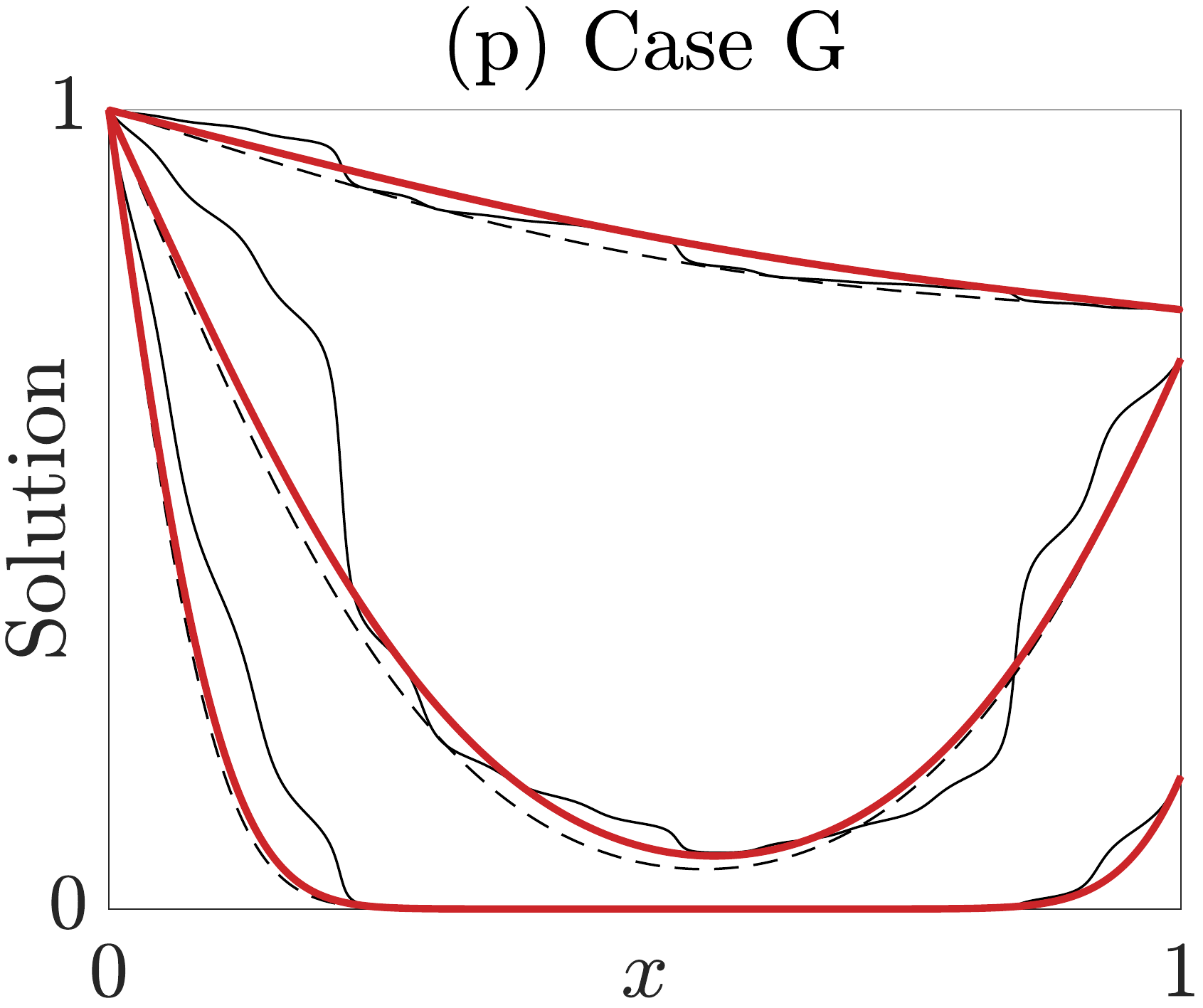}\hspace{0.35cm}\includegraphics[width=\figurewidth]{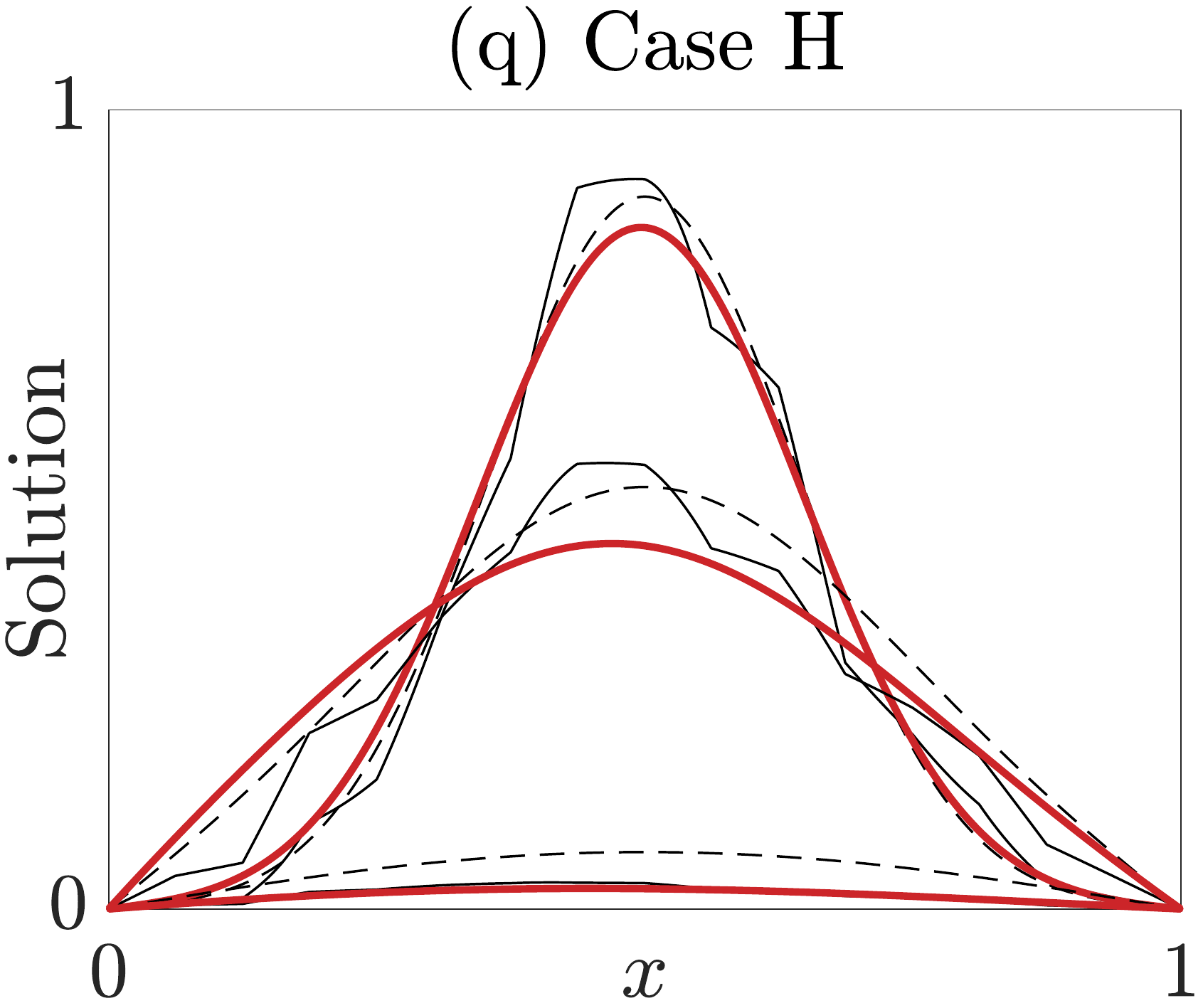}\hspace{0.25cm}
\includegraphics[width=\figurewidth]{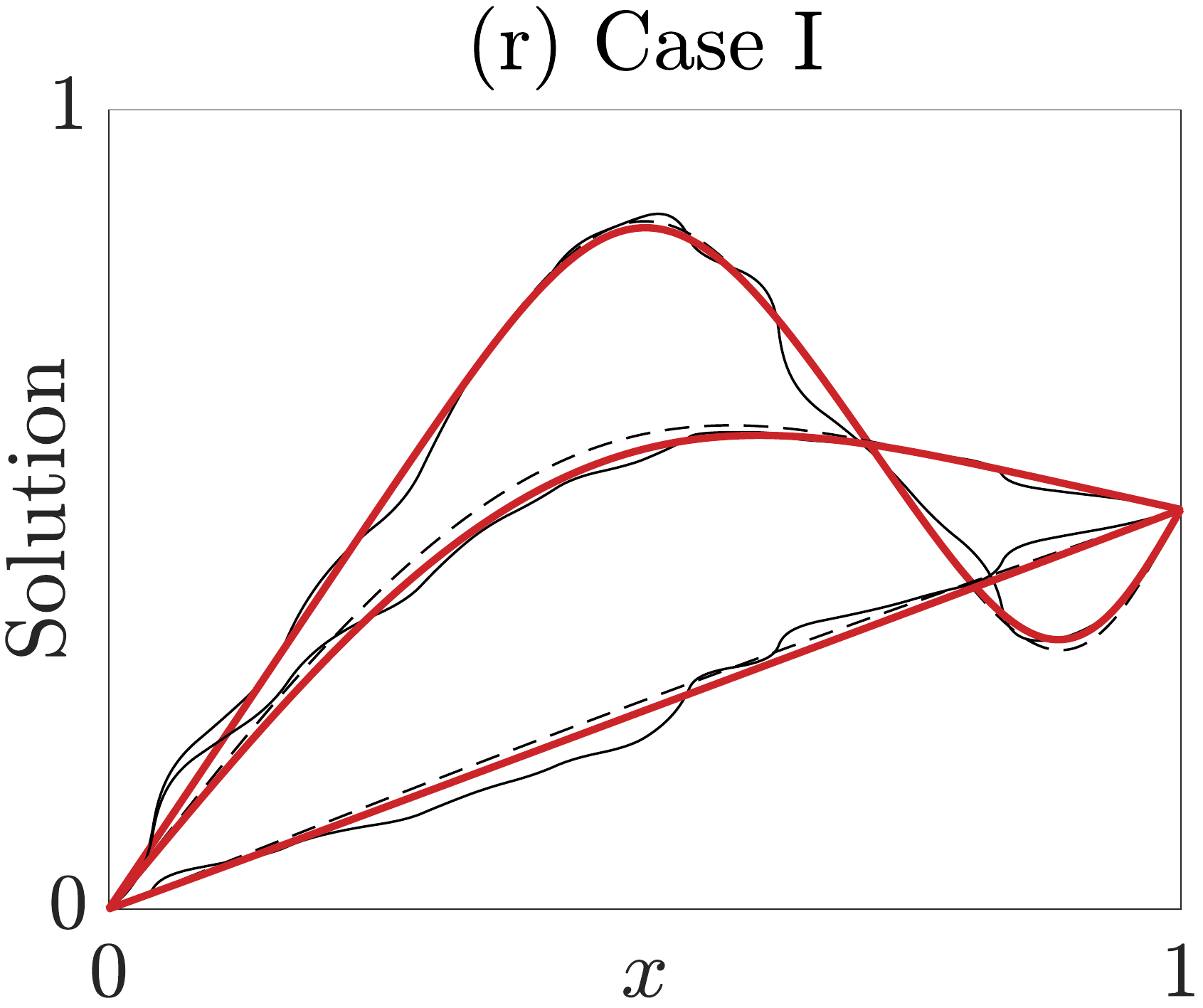}\\
\caption{Results for test cases \revision{A--I} from Table \ref{tab:test_cases} (a)--(i) Diffusivity functions (j)--(r) Solutions of the advection-diffusion homogenized model (\ref{eq:fhm_pde})--(\ref{eq:fhm_bcs}) [$U(x,t)$] and standard diffusion-only homogenized model (\ref{eq:cfhm_pde})--(\ref{eq:cfhm_De}) [$\widetilde{U}(x,t)$] benchmarked against the solution of the heterogeneous model (\ref{eq:hm_pde})--(\ref{eq:hm_bcs}) [$u(x,t)$]. In each of (j)--(r), solutions are shown at three times $t = 10^{-2}$, $10^{-1}$, $1$. The included legend applies to plots (j)--(r) only.}
\label{fig:Results}
\end{figure*}


In our computational experiments, we consider the nine test cases described in Table \ref{tab:test_cases}. In Figure \ref{fig:Results} and Table \ref{tab:errors}, we report results for $N_{x} = 1001$, $N_{t} = 100$ and $\tau = 0.01$. An immediate observation from these results is that the advection-diffusion homogenized model (\ref{eq:fhm_pde})--(\ref{eq:fhm_bcs}) is superior at capturing the smooth behaviour of the heterogeneous model (\ref{eq:hm_pde})--(\ref{eq:hm_bcs}) compared to the diffusion-only homogenized model (\ref{eq:cfhm_pde})--(\ref{eq:cfhm_De}). This is demonstrated in Figures \ref{fig:Results}(j)--(r) by the observation that $U(x,t)$ agrees with $u(x,t)$ better than $\widetilde{U}(x,t)$ does, and in Table \ref{tab:errors} by the smaller values of the mean absolute error for the advection-diffusion homogenized model across all nine test cases.

Several other interesting observations are evident from the individual test cases. Cases A and B involve a constant and linearly decreasing diffusivity perturbed by a sinusoidal function, respectively, and are identical to the two problems shown in Figure \ref{fig:Intro}. For case A, both homogenized models provide a good approximation to the solution behaviour of the heterogeneous model (see Figure \ref{fig:Results}(j)). However, for case B, the advection-diffusion homogenized model significantly outperforms the standard diffusion-only homogenized model. In this test case, the benchmark heterogeneous model produces a solution exhibiting advective behaviour in the positive $x$ direction, which is especially evident from the steady-state solution (see Figure \ref{fig:Results}(k)). This behaviour is accurately captured by the advection-diffusion equation (\ref{eq:fhm_pde}) (note the large value of $\veff$ for case B compared to case A) but cannot be captured by the standard homogenized equation (\ref{eq:cfhm_pde}), where the steady-state solution is linear regardless of the value of $\Defft$ (Figure \ref{fig:Results}(k)). For case B, the new homogenized model also provides a superior match at earlier times (Figure \ref{fig:Results}(k)). Case C is a classical homogenization problem with $D(x)$ expressible as a function of $x/\varepsilon$, where $\varepsilon$ is a small parameter ($\varepsilon = 0.005$ in this case). For this test case, the standard homogenized model (\ref{eq:cfhm_pde})--(\ref{eq:cfhm_De}) is obtained in the homogenization limit $\varepsilon\rightarrow 0$ with $U(x,t)$ and $\widetilde{U}(x,t)$ matching almost precisely with one another (Figure \ref{fig:Results}(l)). \revision{Case C demonstrates the self-averaging property of the diffusion equation where the solutions appear homogeneous when the characteristic length of the heterogeneities is very small relative to the length of the heterogeneous medium \cite{puzko_2019}. The results for cases A, B and C demonstrate that differences between the advection-diffusion homogenized model and the standard diffusion-only homogenized model are accentuated when $D'(x)$ in Eq~(\ref{eq:product_rule}) is large in magnitude relative to $D(x)$. Cases D, E and F further illustrate this observation with large values of $v_{\mathrm{eff}}$ in Table \ref{tab:errors} and pronounced differences between the two homogenized models in Figures \ref{fig:Results}(m)--(o).} Finally, cases \revision{G, H and I} demonstrate that our homogenization approach performs well for time-dependent boundary conditions, non-uniform initial conditions, and non-monotone transitions from initial to steady-state, respectively (see Figure \ref{fig:Results}(p)--(r) and Table \ref{tab:errors}).

\medskip


\section{Conclusions}
\label{sec:conclusion}
Determining an effective homogeneous medium providing the smooth/averaged behaviour of a continuum transport process across a heterogeneous medium is a classical problem in many scientific and engineering disciplines. In this paper, we have presented a homogenization approach for the one-dimensional diffusion equation with spatially variable (heterogeneous) diffusivity. Our approach is noteworthy for the inclusion of an effective advection term in the homogenized equation in addition to the standard effective diffusion term. To identify the effective diffusivity and effective velocity we enforce equality of the spatial average of the steady-state solution of the homogenized and heterogeneous models and equality of the spatial average of a quantity characterising the temporal behaviour of both models. Our proposed homgenization approach requires the solution of two uncoupled boundary value problems Eqs (\ref{eq:hm_pde_ss})--(\ref{eq:hm_bcs_ss}) and (\ref{eq:hm_pde_m})--(\ref{eq:hm_bcs_m2}), over the heterogeneous medium and the solution of a small system of nonlinear equations for the effective parameters (\ref{eq:fhm_nonlinear_system})--(\ref{eq:F2}). Due to the appearance of $g_{0}(t)$ and $g_{L}(t)$ in the boundary value problems, e.g. Eqs (\ref{eq:hm_pde_M})--(\ref{eq:hm_bcs_M2}), the computed effective coefficients depend not only on the spatially-varying diffusivity but also on the boundary conditions imposed in the heterogeneous model.

Computational experiments carried out in Section \ref{sec:computational_experiments} demonstrate, for nine test cases, that the inclusion of an effective advection term leads to an improved approximation of the smooth behaviour of continuum diffusion in a one-dimensional heterogeneous medium. Possible avenues for future work include extension to two or three-dimensional problems or non-Dirichlet boundary conditions. \revision{For higher-dimensional problems, additional constraints are required to identify the additional diffusivities and velocities present in two and three dimensions. One way forward could be to note that the quantities $w(x)$ (\ref{eq:wx}) and $W(x)$ (\ref{eq:Wx}) can be interpreted as the zeroth temporal moment of $s(x) - u(x,t)$ and $S(x) - U(x,t)$, respectively, and introduce constraints on the corresponding higher-order moments \cite{carr_2018a}.} For non-Dirichlet boundary conditions, additional thought is required as the constraint enforcing equality of the spatial average of the steady-state solution (\ref{eq:fhm_constraint1}) provides no information for certain choices of boundary conditions (\ref{eq:hm_bcs}), e.g., for $u(0,t) = 1$ and $\frac{\partial u}{\partial x}(L,t) = 0$, the steady-state solutions, $s(x)$ and $S(x)$, are uniformly equal to one for all $D(x)$, $\Deff$ and $\veff$.

\bibliographystyle{model1-num-names}
\bibliography{references}

\begin{thebibliography}{20}
\expandafter\ifx\csname natexlab\endcsname\relax\def\natexlab#1{#1}\fi
\providecommand{\url}[1]{\texttt{#1}}
\providecommand{\href}[2]{#2}
\providecommand{\path}[1]{#1}
\providecommand{\DOIprefix}{doi:}
\providecommand{\ArXivprefix}{arXiv:}
\providecommand{\URLprefix}{URL: }
\providecommand{\Pubmedprefix}{pmid:}
\providecommand{\doi}[1]{\href{http://dx.doi.org/#1}{\path{#1}}}
\providecommand{\Pubmed}[1]{\href{pmid:#1}{\path{#1}}}
\providecommand{\bibinfo}[2]{#2}
\ifx\xfnm\relax \def\xfnm[#1]{\unskip,\space#1}\fi
\bibitem[{Chen and Ren(2008)}]{chen_2008}
\bibinfo{author}{F.~Chen}, \bibinfo{author}{L.~Ren},
\newblock \bibinfo{title}{Application of the finite difference heterogeneous
  multiscale method to the {Richards’} equation},
\newblock \bibinfo{journal}{Water Resour. Res.} \bibinfo{volume}{44}
  (\bibinfo{year}{2008}) \bibinfo{pages}{W07413}.
\bibitem[{Perr{\'e} and Turner(2002)}]{perre_2002}
\bibinfo{author}{P.~Perr{\'e}}, \bibinfo{author}{I.~W. Turner},
\newblock \bibinfo{title}{A heterogeneous wood drying computational model that
  accounts for material property variation across growth rings},
\newblock \bibinfo{journal}{Chem. Eng. J.} \bibinfo{volume}{86}
  (\bibinfo{year}{2002}) \bibinfo{pages}{117--131}.
\bibitem[{Matzavinos and Ptashnyk(2016)}]{matzavinos_2016}
\bibinfo{author}{A.~Matzavinos}, \bibinfo{author}{M.~Ptashnyk},
\newblock \bibinfo{title}{Homogenization of oxygen transport in biological
  tissues},
\newblock \bibinfo{journal}{Appl. Anal.} \bibinfo{volume}{95}
  (\bibinfo{year}{2016}) \bibinfo{pages}{1013--1049}.
\bibitem[{Huysmans and Dassargues(2007)}]{huysmans_2007}
\bibinfo{author}{M.~Huysmans}, \bibinfo{author}{A.~Dassargues},
\newblock \bibinfo{title}{Equivalent diffusion coefficient and equivalent
  diffusion accessible porosity of a stratified porous medium},
\newblock \bibinfo{journal}{Transport Porous Med.} \bibinfo{volume}{66}
  (\bibinfo{year}{2007}) \bibinfo{pages}{421--438}.
\bibitem[{Auriault(1991)}]{auriault_1991}
\bibinfo{author}{J.~L. Auriault},
\newblock \bibinfo{title}{Heterogeneous medium. {Is} an equivalent macroscopic
  description possible?},
\newblock \bibinfo{journal}{Int. J. Eng. Sci.} \bibinfo{volume}{29}
  (\bibinfo{year}{1991}) \bibinfo{pages}{785--795}.
\bibitem[{Roberts(2010)}]{roberts_2010}
\bibinfo{author}{A.~J. Roberts},
\newblock \bibinfo{title}{The harmonic mean renormalises random diffusion
  across a spatial multigrid},
\newblock \bibinfo{journal}{ANZIAM J.} \bibinfo{volume}{51}
  (\bibinfo{year}{2010}) \bibinfo{pages}{C83--C96}.
\bibitem[{Davit et~al.(2013)Davit, Bell, Byrne, Chapman, Kimpton, Lang,
  Leonard, Oliver, Pearson, Shipley, Waters, Whiteley, Wood, and
  Quintard}]{davit_2013}
\bibinfo{author}{Y.~Davit}, \bibinfo{author}{C.~G. Bell},
  \bibinfo{author}{H.~M. Byrne}, \bibinfo{author}{L.~Chapman},
  \bibinfo{author}{L.~Kimpton}, \bibinfo{author}{G.~Lang},
  \bibinfo{author}{K.~Leonard}, \bibinfo{author}{J.~Oliver},
  \bibinfo{author}{N.~Pearson}, \bibinfo{author}{R.~Shipley},
  \bibinfo{author}{S.~Waters}, \bibinfo{author}{J.~Whiteley},
  \bibinfo{author}{B.~Wood}, \bibinfo{author}{M.~Quintard},
\newblock \bibinfo{title}{Homogenization via formal multiscale asymptotics and
  volume averaging: how do the two techniques compare?},
\newblock \bibinfo{journal}{Adv. Water Resour.} \bibinfo{volume}{62}
  (\bibinfo{year}{2013}) \bibinfo{pages}{178--206}.
\bibitem[{Carr et~al.(2017)Carr, Turner, and Perr{\'e}}]{carr_2017a}
\bibinfo{author}{E.~J. Carr}, \bibinfo{author}{I.~W. Turner},
  \bibinfo{author}{P.~Perr{\'e}},
\newblock \bibinfo{title}{Macroscale modelling of multilayer diffusion: Using
  volume averaging to correct the boundary conditions},
\newblock \bibinfo{journal}{Appl. Math. Model.} \bibinfo{volume}{47}
  (\bibinfo{year}{2017}) \bibinfo{pages}{600--618}.
\bibitem[{Abdulle and E(2003)}]{abdulle_2003}
\bibinfo{author}{A.~Abdulle}, \bibinfo{author}{W.~E},
\newblock \bibinfo{title}{Finite difference heterogeneous multi-scale method
  for homogenization problems},
\newblock \bibinfo{journal}{J. Comput. Phys.} \bibinfo{volume}{191}
  (\bibinfo{year}{2003}) \bibinfo{pages}{18--39}.
\bibitem[{Samaey et~al.(2005)Samaey, Roose, and Kevrekidis}]{samaey_2005}
\bibinfo{author}{G.~Samaey}, \bibinfo{author}{D.~Roose}, \bibinfo{author}{I.~G.
  Kevrekidis},
\newblock \bibinfo{title}{The gap-tooth scheme for homogenization problems},
\newblock \bibinfo{journal}{Multiscale Model. Sim.} \bibinfo{volume}{4}
  (\bibinfo{year}{2005}) \bibinfo{pages}{278--306}.
\bibitem[{Crank(1975)}]{crank_1975}
\bibinfo{author}{J.~Crank}, \bibinfo{title}{The mathematics of diffusion},
  \bibinfo{edition}{2nd} ed., \bibinfo{publisher}{Oxford University Press},
  \bibinfo{year}{1975}.
\bibitem[{Hornung(1997)}]{hornung_1997}
\bibinfo{author}{U.~Hornung}, \bibinfo{title}{Homogenization and Porous Media},
  \bibinfo{publisher}{Springer-Verlag}, \bibinfo{address}{New York},
  \bibinfo{year}{1997}.
\bibitem[{Holmes(2013)}]{holmes_2013}
\bibinfo{author}{M.~H. Holmes}, \bibinfo{title}{Introduction to Perturbation
  Methods}, \bibinfo{edition}{2nd} ed., \bibinfo{publisher}{Springer},
  \bibinfo{address}{New York}, \bibinfo{year}{2013}.
\bibitem[{Pavliotis and Stuart(2008)}]{pavliotis_2008}
\bibinfo{author}{G.~A. Pavliotis}, \bibinfo{author}{A.~M. Stuart},
  \bibinfo{title}{Multiscale Methods: Averaging and Homogenization},
  \bibinfo{publisher}{Springer}, \bibinfo{address}{New York},
  \bibinfo{year}{2008}.
\bibitem[{Ray et~al.(2018)Ray, Rupp, Schulz, and Knabner}]{ray_2018}
\bibinfo{author}{N.~Ray}, \bibinfo{author}{A.~Rupp},
  \bibinfo{author}{R.~Schulz}, \bibinfo{author}{P.~Knabner},
\newblock \bibinfo{title}{Old and new approaches predicting the diffusion in
  porous media},
\newblock \bibinfo{journal}{Transport Porous Med.} \bibinfo{volume}{124}
  (\bibinfo{year}{2018}) \bibinfo{pages}{803--824}.
\bibitem[{Carr and Simpson(2019)}]{carr_2019a}
\bibinfo{author}{E.~J. Carr}, \bibinfo{author}{M.~J. Simpson},
\newblock \bibinfo{title}{New homogenization approaches for stochastic
  transport through heterogeneous media},
\newblock \bibinfo{journal}{J. Chem. Phys.} \bibinfo{volume}{150}
  (\bibinfo{year}{2019}) \bibinfo{pages}{044104}.
\bibitem[{Carr(2019)}]{carr_2019b}
\bibinfo{author}{E.~J. Carr},
\newblock \bibinfo{title}{Rear-surface integral method for calculating thermal
  diffusivity from laser flash experiments},
\newblock \bibinfo{journal}{Chem. Eng. Sci.} \bibinfo{volume}{199}
  (\bibinfo{year}{2019}) \bibinfo{pages}{546--551}.
\bibitem[{Carr and Simpson(2018)}]{carr_2018a}
\bibinfo{author}{E.~J. Carr}, \bibinfo{author}{M.~J. Simpson},
\newblock \bibinfo{title}{Accurate and efficient calculation of response times
  for groundwater flow},
\newblock \bibinfo{journal}{J. Hydrol.} \bibinfo{volume}{558}
  (\bibinfo{year}{2018}) \bibinfo{pages}{470--481}.
\bibitem[{Ellery et~al.(2012)Ellery, Simpson, McCue, and Baker}]{ellery_2012}
\bibinfo{author}{A.~J. Ellery}, \bibinfo{author}{M.~J. Simpson},
  \bibinfo{author}{S.~W. McCue}, \bibinfo{author}{R.~E. Baker},
\newblock \bibinfo{title}{Critical time scales for advection-diffusion-reaction
  processes},
\newblock \bibinfo{journal}{Phys. Rev. E} \bibinfo{volume}{85}
  (\bibinfo{year}{2012}) \bibinfo{pages}{041135}.
\bibitem[{Puzko and Merzlikin(2019)}]{puzko_2019}
\bibinfo{author}{R.~S. Puzko}, \bibinfo{author}{A.~M. Merzlikin},
\newblock \bibinfo{title}{Homogenization of maxwell’s equations in layered
  system beyond static approximation}  (\bibinfo{year}{2019})
  \bibinfo{pages}{\href{https://arxiv.org/abs/1708.01661}{arXiv:1708.01661}}.

\end{thebibliography}

\end{document}